\documentstyle[epsfig,psfig]{texas}

\newcommand{\ddd}{{\mathrm d}}
\newcommand{\Hconf}{{\mathcal H}}

\newcommand{\ETC}{etc}

\newcommand{\IE}{i.e.}
\newcommand{\EG}{e.g.}
\newcommand{\ETAL}{{\it et al.}}

\newcommand{\EQ}[1]{eq.~(#1)}
\newcommand{\EEQ}[1]{Eq.~(#1)}
\newcommand{\EQNS}[1]{eqns.~(#1)}
\newcommand{\EEQNS}[1]{Eqns.~(#1)}
\newcommand{\FIG}[1]{fig.~{#1}}
\newcommand{\FFIG}[1]{Fig.~{#1}}

\begin{document}

\title{Cosmic microwave background anisotropies seeded by coherent
topological defects~: a semi-analytic approach}

\author{Jean-Philippe Uzan ${}^1$, Nathalie Deruelle $^{1,2}$ and 
Alain Riazuelo$^1$\\}

\address{
(1) D\'epartement d'Astrophysique Relativiste et de Cosmologie, \\
UPR 176 du Centre National de la Recherche Scientifique, Observatoire
de Paris, 92195 Meudon, France}
\address{
(2) DAMTP, University of Cambridge, \\ Silver Street, Cambridge, CB3
9EW, England \\
\rm Email: \\
uzan@amorgos.unige.ch, Nathalie.Deruelle@obspm.fr,
Alain.Riazuelo@obspm.fr}

\begin{abstract}
We consider a perfectly homogeneous, isotropic and spatially flat
universe which undergoes a sudden phase transition producing
topological defects. We assume that these defects form a coherent
network which scales like the background density during the radiation
and matter dominated eras, and describe them in terms of a few free
functions. We carefully model the loss of scaling invariance during
the transition from radiation to matter dominated era. We choose a
simple set for the free functions, compute the microwave background
temperature anisotropies generated by such a network, and compare our
results to previous calculations.
\end{abstract}

\section{Introduction} 
\label{sec_intro}

The origin of the small inhomogeneities in the cosmic fluids which
eventually evolved into the large scale structures that we observe
today in the universe is still under debate. Two classes of models are
at present in competition. The inflationary models on one hand
\cite{infl_guth} explain those inhomogeneities by the amplification of
quantum fluctuations at the end of an accelerated phase of expansion
of the universe \cite{infl_gen}. On the other hand topological defects
appearing during a phase transition in the early universe
\cite{kibble} can also seed inhomogeneities \cite{def_gen}.

A way to discriminate between those two classes of models is to
compute the cosmic microwave background (CMB) anisotropies they
predict and confront the results to observations.  At present the
anisotropies on large angular scales are known from the COBE
measurements \cite{cobe_res}. On smaller scales, the present balloon
and ground experiments \cite{rev_exp} are still unprecise but the
measurements should improve when the MAP \cite{map} and Planck
\cite{plank} satellite missions are launched.

On the theoretical front, the CMB anisotropy predictions from the
inflationary models (at least the simplest ones) are robust, and can
now in many cases be obtained by plugging into the CMBFAST code
\cite{cmbfast_art}\cite{cmbfast_code} the initial conditions given by
the particular model at hand.  On the other hand the predictions from
defect scenarios are still unclear, the reason being that the defects,
which act as a continuous source of inhomogeneities, are difficult to
model. Indeed the description of the formation, evolution and decay of
defects requires heavy numerical simulations (see \EG{}
\cite{def_num1}-\cite{def_num5}), and the CMB anisotropies their
networks create seem to be very sensitive to their detailed
structure. Recent results \cite{abr}\cite{pst1} indicate that defect
model predictions do not agree with the data. It is clear however that
more work needs to be done before rejecting topological defects as a
possible explanation for the CMB anisotropies as they may for example
have a more complex internal structure (see \EG{} \cite{patrick}) than
that assumed in those numerical models.

Considering the complexity of modeling realistic topological defects,
the semi-analytic approach initiated by Durrer and collaborators
\cite{durrer} seems to be a fruitful compromise. The idea is to model
the defect network by a stress-energy tensor which acts as a source to
the linearized Einstein equations and induces inhomogeneities in the
cosmic fluids. The ten components of this stress-energy tensor are
then drastically constrained by a number of physical requirements.  In
this paper, we shall impose, as in \cite{durrer}\cite{pst2}, that the
defect network
\begin{enumerate}
  \item \label{enum1} is statistically homogeneous and isotropic
  (hence obeys the copernican principle),

  \item \label{enum2} is created at a phase transition in an up to
  then perfectly homogeneous and isotropic universe (hence obeys
  specific causality \cite{turok} and matching \cite{dlu} conditions),

  \item \label{enum3} evolves deep in the radiation era and deep in
  the matter era in a way which is statistically independent of time
  (scaling requirement \cite{durrer}).

We shall also make the further assumptions that the network

  \item \label{enum4} satisfies the conservation equations,

  \item \label{enum5} is statistically coherent (see \EG{} \cite{dlu}
  and below for a precise definition).
\end{enumerate}
Of all those requirements the latter two are stringent and probably
not fulfilled by realistic defect networks (see \EG{}
\cite{pst1}\cite{durrer}\cite{chm}).

Now, as for assumption (\ref{enum5}), it is a sensible one to make, as
incoherent sources can in principle be described as a sum of coherent
ones \cite{pst1} (see \S \ref{ssec_coh}). As for assumption
(\ref{enum4}), all authors who describe semi-analytically the defects
as a fluid have up to now made it. All authors who consider global
defects construct {\it conserved} stress energy tensors (see \EG{}
\cite{turok2}). Those who study local strings may have to add an extra
fluid so that the total stress energy tensor is conserved (see \EG{}
\cite{abr}\cite{turok2}). Only Contaldi \ETAL{} \cite{chm}, as far as
we are aware, have questionned it.  In order to take into account the
network loss of energy through gravitational radiation or particle
production, they have added an extra fluid which compensates for the
non conservation of the defect stress-energy tensor (which comes from
excising small loops from the numerical simulation in order for the
network to scale).  The problem with such a modelling (as Contaldi
\ETAL{} themselves note) is, first, that the results are very
sensitive to the equation of state of that extra fluid (which is left
as a free parameter), and, second, that the energy lost by the network
is transferred exclusively to that extra fluid and not to the
background photons, baryons, Cold Dark Matter (CDM) and
neutrinos. Therefore the modelling of the defect network energy loss
is still in a primitive stage and needs to be improved by a careful
analysis of the microphysics involved in the evolution of the network
(which includes friction effects, particle and gravitational radiation
production by decaying loops, as well as electromagnetic interactions
with the plasma when the network carries currents). In this paper, in
order to keep arbitrariness to a minimum, we choose to ignore the
network energy losses.

Those five conditions being imposed, the stress-energy tensor of the
defects will turn out to depend only on four free functions, two
describing its scalar part and two describing respectively its
vectorial and tensorial parts. Moreover, for a particular ansatz which
garantees that the defects evolve from the two scaling regimes in such
a way that the conservation equations are satisfied, those four
functions will, as we shall see, depend only on one variable $u$ and
tend to constants when $u\to0$, and to $0$ when $u\to\infty$.

We shall choose simple ans\"atze for those four functions (their
precise shape can in principle be obtained by numerical simulations)
and solve the well-known (see \EG{} \cite{pert_gen2}) linearised
Einstein equations which couple the defect network to the cosmic fluid
inhomogeneities. Finally we shall compute numerically the CMB
anisotropies those inhomogeneities produce.

In agreement with \cite{durrer} (as well as with
\cite{abr}\cite{pst2}) we shall see that those anisotropies are very
sensitive to the choices made for the free functions describing the
defects.  Our results therefore confirm previous work but the loss of
scaling invariance during the transition is treated more
carefully. They are also more complete as they include the vector and
tensor contributions.

The paper is organised as follows~: in \S \ref{sec2} we write down the
linearised Einstein equations (in longitudinal gauge) describing the
evolution of the cosmic fluid inhomogeneities driven by the defect
stress-energy tensor (we suppose that the material content of the
universe is a mixture of four perfect fluids~: CDM, neutrinos, baryons
and photons, the latter two fluids being tightly coupled before
decoupling). In \S \ref{sec_stress} we see in a detailed way how the
five conditions given above constrain the defect stress-energy
tensor. In \S \ref{sec_ic} we solve analytically the linearised Einstein
equations deep in the radiation era, when all scales of interest today
were larger than the Hubble radius, and we set the initial conditions
on the cosmic fluid perturbations, following \cite{dlu}. In \S
\ref{sec_cmb} we give the correlation function of the CMB anisotropies in
function of the cosmic fluid perturbations, assuming instantaneous
decoupling (our computation is slightly different from the standard
one --- see \EG{} \cite{pert_gen}). Finally in \S \ref{sec_res} we
compute numerically the multipole coefficients $C_\ell$ of the CMB
anisotropy correlation function using a perfect fluid code and compare
our results to previous computations. We discuss the approximations
made, to wit perfect fluid description, tight coupling and
instantaneous decoupling hypotheses, by comparing our perfect fluid
code to a Boltzmann one. We also show how to evaluate analytically the
vector and tensor contributions to the $C_\ell$ for small $\ell$.

\section{Einstein equations}
\label{sec2}

\subsection{The background}

The universe at large appears to be remarkably homogeneous and
isotropic and governed by the gravitational force created by its
material content. We suppose that that was always so, and describe it
in a first approximation by a Robertson-Walker geometry (that we shall
for simplicity suppose spatially flat) whose time evolution satisfies
Friedmann's equations. We suppose that its material content is a
mixture of perfect fluids made of decoupled black body radiations of
photons and massless neutrinos, of baryons with which the photons were
coupled until their temperature dropped to $\approx 3000$ K, and of
uncoupled ``Cold Dark Matter'' (CDM).  We thus take the line element
to be~:
\begin{equation}
\ddd s^2=a^2(\eta)(-\ddd\eta^2+\delta_{ij}\ddd x^i \ddd x^j),
\end{equation} 
where $\eta$ is conformal time (we set the velocity of light equal to
1), where $x^i$, $i=1,2,3$ are three cartesian coordinates and
$\delta_{ij}$ is the Kronecker symbol, and where $a(\eta)$ is the
scale factor.

The conservation equations for the fluids read~:
\begin{equation}
\label{cons_bg_fluid}
\rho'_n=-3{\mathcal H}(\rho_n+P_n),
\end{equation}
where $\rho_n$ and $P_n$ are the energy density and pressure of the
$n$-th fluid and where ${\mathcal H}\equiv a'/a$ is the comoving
Hubble radius, with a prime denoting a derivative with respect to
conformal time. The radiation fluids of photons and neutrinos both
have a pressure equal to a third of their energy density, and the
fluids of baryons and CDM have zero pressure (whence they have become
non-relativistic). Hence the two former scale as $a^{-4}$ and the two
latter as $a^{-3}$.

The Friedmann equation is~:
\begin{equation}
{\mathcal H}^2={\kappa \frac{a^2}{3}}\rho_t,
\label{friedmann}
\end{equation}
where $\kappa\equiv8\pi {\mathcal G}$ is Einstein's constant and
$\rho_t\equiv\sum_n\rho_n$.

Combining \EQNS{\ref{cons_bg_fluid}} and (\ref{friedmann}) yields~:
\begin{equation}
\label{h2}
h^2=\frac{1}{x^2}(\Omega_r^0+x\Omega_m^0),
\end{equation}
with
\begin{eqnarray}
\Omega_r^0 & \equiv & \Omega_\gamma^0+\Omega_\nu^0, \\
\Omega_m^0 & \equiv & \Omega_b^0+\Omega_c^0, \\
x & \equiv & a/a_0 \\ h & \equiv & {\mathcal H}/{\mathcal H}_0,
\end{eqnarray}
where $\Omega_n\equiv\rho_n/\rho_t$ (with $\sum_n\Omega_n=1$), where
the index zero means the present time and where the indices
$\gamma,\nu,b,c$ refer respectively to the photons, neutrinos, baryons
and CDM.

The numerical values of $\Omega_n^0$ are known from~:
\begin{enumerate}
  \item the value of the Hubble constant $H_0\equiv {\mathcal
  H}_0/a_0$, which determines $\rho_t^0$ via \EQ{\ref{friedmann}} (we
  shall take $H_0=100h_0$ km sec$^{-1}$Mpc$^{-1}$ with $h_0=0.5$),

  \item the temperature of the microwave background ($2.726$ K), which
  determines $\rho_\gamma^0$ via Stefan's law,

  \item electroweak theory, which determines the ratio
  $\rho_\nu^0/\rho_\gamma^0=0.68132$ (assuming three families of
  massless, non-degenerate neutrinos),

  \item nucleosynthesis, which determines $\rho_b^0$ (we shall take
  $\Omega_b^0h_0^2=0.0125$).
\end{enumerate}
All that yields~:
\begin{eqnarray}
\Omega_\gamma^0 & \simeq & 9.4\times10^{-5},  \\
\Omega_\nu^0    & \simeq & 6.3\times10^{-5},  \\
\Omega_b^0      & =      & 5.0\times 10^{-2}, \\
\Omega_c^0      & \simeq & 0.95.
\end{eqnarray}
The value $x_{\rm d}$ of $x$ at decoupling is given by the ratio of the
present temperature of the photon background and its temperature at
decoupling~: $x_{\rm d}\approx 0.9\times10^{-3}$.  Equality of matter and
radiation is defined by $x_{\rm eq}\equiv\Omega_r^0/\Omega_m^0$ and
occurs before decoupling.

Equation (\ref{h2}) can be easily solved to give $x(\eta)$~:
\begin{equation}
\label{eta}
\eta_0-\eta=\frac{2}{\Omega_m^0{\mathcal
H}_0}\left[1-\sqrt{\Omega_r^0+x\Omega_m^0}\right].
\end{equation}
In the radiation era ($x\ll x_{\rm eq}$), $x\propto\eta$, in the
matter era ($x\gg x_{\rm eq}$), $x\propto \eta^2$ and in both eras
${\mathcal H}\propto 1/\eta$.

Finally, when the scale factor behaves as $a(\eta)\propto\eta^q$ with
$q>0$ when $\eta\to0$ (which is the case in the standard scenario we
consider here) the particle horizon of the point $x^i$ is defined by
$r=\eta$ (where $r\equiv|x^i-x'^i|$ is the comoving spatial distance
from $x^i$) and is proportional to the comoving Hubble radius
${\mathcal H}^{-1}$.

\subsection{Perturbation equations}

Following Bardeen \cite{bardeen} (see also \cite{stewart}) we split
the perturbations of the geometry and the matter variables into
(spatial) scalar, vector and tensor components (for reviews of this
formalism, see \EG{} \cite{pert_gen2}). We shall work in the
longitudinal (also called newtonian) gauge in which the line element
of a perturbed Friedmann-Robertson-Walker space time reads~:
\begin{equation}
\label{frw_pert}
\ddd s^2=a^2(\eta)\{-(1+2\Phi)\ddd\eta^2-2\bar \Phi_i
\ddd x^i\ddd\eta+[(1-2\Psi)\delta_{ij}+2\bar E_{ij}]
\ddd x^i \ddd x^j\}.
\end{equation}
Here and in the following all barred spatial vectors (such as
$\bar\Phi_i$) are divergenceless ($\partial_i\bar\Phi^i=0$), all
barred spatial tensors (such as $\bar E_{ij}$) are divergenceless and
traceless ($\partial_i\bar E^{ij}=0, \bar E^i_i=0$) and all spatial
indices are moved with the Kronecker symbol $\delta_{ij}$. The six
functions $(\Phi, \Psi)$, $\bar\Phi_i$ and $\bar E_{ij}$ are
respectively the scalar, vector and tensor parts of the metric
perturbations in longitudinal gauge.

We shall treat the material content of the universe as a mixture of
{\sl perfect} fluids (see \S \ref{sec_res} below for a discussion of this
hypothesis). The perturbations of the stress-energy tensor of the
$n$-th perfect fluid in this perturbed universe are~:
\begin{eqnarray}
\delta T^n_{00} & = & 
  a^2\rho_n(\delta_n+2\Phi) \\
\delta T^n_{0i} & = & 
  a^2[\rho_n\bar \Phi_i-(\bar v_{i, n}+\partial_i v_n)(\rho_n+P_n)] \\
\delta T^n_{ij} & = & 
  a^2 P_n[2\bar E_{ij}+\delta_{ij}(\delta_n-2\Psi)]
\end{eqnarray} 
where $\delta_n\equiv\delta\rho_n/\rho_n$ is its density contrast and
$\partial_i v_n+\bar v_{i, n}$ its velocity perturbation.

The stress-energy tensor of the topological defects is in itself a
small perturbation (this is the so-called ``stiff approximation'', see
\EG{} \cite{veersteb}). We decompose its components as~:
\begin{eqnarray}
\label{svt_00}
\frac{\kappa}{{\mathcal H}_0^2}\Theta_{00}
 & = & \rho^s, \\
\label{svt_0i}
\frac{\kappa}{{\mathcal H}_0^2}\Theta_{0i}
 & = & -(\bar v_i^s+\partial_i v^s), \\
\label{svt_ij}
\frac{\kappa}{{\mathcal H}_0^2}\Theta_{ij}
 & = &  \bar\Pi^s_{ij}+\partial_i\bar\Pi^s_j+\partial_j\bar\Pi^s_i
       +\delta_{ij}\left(P^s-\frac{1}{3}\Delta\Pi^s\right)
       +\partial_{ij}\Pi^s.
\end{eqnarray}

The ten source functions $(\rho^s, P^s, v^s, \Pi^s), (\bar
v^s_i,\bar\Pi^s_i)$ and $\bar\Pi^s_{ij}$ will be discussed later.

We shall write the evolution equations in Fourier space, the Fourier
transform of any function $f(x^i,\eta)$ being defined (formally) as~:
\begin{eqnarray}
\label{TF}
 \hat f(k^i,\eta) & = & 
\frac{{\mathcal H}_0^3}{(2\pi)^\frac{3}{2}}
\int \ddd^3x e^{-{\rm i}k_ix^i}f(x^i,\eta), \\
f(x^i,\eta) & = & 
\frac{1}{(2\pi)^\frac{3}{2} {\mathcal H}_0^3}
\int \ddd^3 k e^{{\rm i}k_ix^i}\hat f(k^i,\eta).
\end{eqnarray}
(The factor ${\mathcal H}_0^3$ insures that $\hat f$ and $f$ have the
same dimension.)

As announced in the introduction, we assume that the defects interact
only gravitationally with the rest of the matter. Therefore their
stress-energy tensor is conserved, which yields for all (but the
$k^i=0$) modes~:
\begin{eqnarray}
\label{def_cons_s0}
\frac{\ddd \hat\rho^s}{\ddd \eta} & = &
 -\Hconf\hat\rho^s-3\Hconf\hat P^s +k\hat V^s, \\
\label{def_cons_si}
\frac{\ddd\hat V^s}{\ddd \eta} & = & 
  -2\Hconf\hat V^s 
  +k\left( -\hat P^s+\frac{2}{3}\hat\pi^s\right), \\ 
\label{def_cons_vi}
\frac{\ddd\hat{\bar v}_i^s}{\ddd \eta} & = & 
  -2\Hconf \hat{\bar v}_i^s + k\hat{\bar\pi}^s_i,
\end{eqnarray}
where we have introduced $k\equiv\sqrt{k_ik^i}$, where a barred vector
is orthogonal to $k^i$ and where
\begin{eqnarray}
\hat V^s          & \equiv & k\hat v^s, \\
\hat\pi^s         & \equiv & k^2\hat\Pi^s, \\
\hat{\bar\pi}^s_i & \equiv & k\hat{\bar\Pi}^s_i.
\end{eqnarray}

After decoupling the fluids of neutrinos, photons, baryons and CDM are
decoupled. Their stress-energy tensors are therefore separately
conserved, which yields~: 
\begin{eqnarray}
\label{pert_nu}
\frac{\ddd\hat\delta^\flat_\nu}{\ddd \eta} = \frac{4}{3}k\hat V_\nu
\quad,\quad 
\frac{\ddd\hat V_\nu}{\ddd \eta}              
 & = & -k\left(\hat\Phi+\frac{1}{4}\hat\delta_\nu\right)
\quad,\quad 
\frac{\ddd\hat{\bar v}^i_\nu}{\ddd \eta} =  0, \\
\label{pert_phot_ad}
\frac{\ddd\hat\delta^\flat_\gamma}{\ddd \eta}
   = \frac{4}{3}k\hat V_\gamma
\quad,\quad 
\frac{\ddd\hat V_\gamma}{\ddd \eta}           
 & = & -k\left(\hat\Phi+\frac{1}{4}\hat\delta_\gamma\right)
\quad,\quad 
\frac{\ddd\hat{\bar v}^i_\gamma}{\ddd \eta} = 0, \\
\label{pert_bar_apd}
\frac{\ddd\hat\delta^\flat_b}{\ddd \eta} = k\hat V_b
\quad,\quad 
\frac{\ddd \hat V_b}{\ddd \eta} & = & -\Hconf\hat V_b -k\hat\Phi
\quad,\quad 
\frac{\ddd \hat{\bar v}^i_b}{\ddd x} = -\Hconf \hat{\bar v}^i_b, \\
\label{pert_cdm}
\frac{\ddd\hat\delta^\flat_c}{\ddd \eta} = k\hat V_c
\quad,\quad 
\frac{\ddd \hat V_c}{\ddd \eta} & = & -\Hconf\hat V_c -k\hat\Phi
\quad,\quad 
\frac{\ddd \hat{\bar v}^i_c}{\ddd \eta} = -\Hconf \hat{\bar v}^i_c.
\end{eqnarray}
where we have introduced
\begin{eqnarray}
\hat\delta^\flat_n & \equiv & 
  \hat\delta_n-3\left(1+\frac{P_n}{\rho_n}\right)\hat\Psi, \\
\hat V_n           & \equiv & k\hat v_n.
\end{eqnarray}
Before decoupling the density contrasts and velocity perturbations of
the neutrino and CDM energy fluids still obey \EQNS{\ref{pert_nu}} and
(\ref{pert_cdm}) but the baryons and photons being tightly coupled,
\EQNS{\ref{pert_phot_ad}} and (\ref{pert_bar_apd}) must be replaced by
(using also the fact that the number of baryons is conserved)
\begin{eqnarray}
\label{pert_bar_avd_scal}
\frac{\ddd\hat\delta^\flat_b}{\ddd \eta} = k\hat V_b
\quad&,&\quad 
\frac{\ddd\hat\delta^\flat_\gamma}{\ddd \eta} = \frac{4}{3}k\hat V_\gamma, \\
\label{pert_phot_avd_scal}
\hat V_b = \hat V_\gamma
\quad&,&\quad 
\frac{\ddd\hat V_\gamma}{ \ddd \eta}
 = -\Hconf R\hat V_\gamma 
       -k  \left(\frac{1}{4}(1-R)\hat\delta_\gamma+\hat\Phi\right), \\
\label{pert_pb_avd_vec}
\hat{\bar v}_{i, b} = \hat{\bar v}_{i, \gamma}
\quad&,&\quad  
\frac{\ddd\hat{\bar v}_{i, \gamma}}{ \ddd \eta}
 = -\Hconf R \hat{\bar v}_{i, \gamma},
\end{eqnarray}
where we have introduced $R \equiv (3\alpha x)/(4+3\alpha x)$,
$\alpha\equiv\Omega_b^0/\Omega_\gamma^0$. (It is clear that imposing
that the density contrasts and velocity perturbations of the baryon
and photon fluids obey
\EQNS{\ref{pert_bar_avd_scal}-\ref{pert_pb_avd_vec}} until decoupling and
(\ref{pert_phot_ad},\ref{pert_bar_apd}) after decoupling, all
variables being continuous at $x=x_{\rm d}$, is a first approximation~: see
\S \ref{sec_res} below.)

The Einstein equations give, after linearisation, the scalar and
vector metric perturbations algebraically in function of the matter
perturbations as
\begin{eqnarray}
\label{poisson}
\hat\Psi & = &
\frac{-1}{k^2 + 3 (\Hconf^2-\ddd \Hconf / \ddd\eta)}\times \\
\nonumber & & 
\left[3\Hconf^2\left(\sum_n{\Omega_n\left(\hat\delta^\flat_n
                            -3(1+\omega_n)\frac{\Hconf}{k}\hat V_n\right)}
               \right)
      + \Hconf_0^2\left(\hat\rho^s-3\frac{\Hconf}{k}\hat V^s\right)\right]
, \\
\label{poisson2}
\hat\Phi & = & \hat \Psi-\frac{\hat\pi^s}{q^2}, \\
\label{poisson_vec}
\hat{\bar\Phi}_i & = & \frac{2}{k^2} \left(
3\Hconf^2 \sum_n{\Omega_n(1+\omega_n)\hat{\bar v}_{i, n}}
+ \Hconf_0^2 \hat{\bar v}_i^s \right).
\end{eqnarray}
(The other Einstein equations are redundant because of Bianchi's
identities. The index $n$ runs on the four background fluids, and we
have set $\omega_b = \omega_c = 0$, $\omega_\nu = \omega_\gamma = 1/3$
and $q \equiv k/\Hconf_0$.) Finally the tensor metric perturbations
are not determined algebraically~: they solve the differential
equation
\begin{equation}
\label{pert_tens}
\frac{\ddd^2\hat{\bar E}_{ij}}{\ddd \eta^2}-2\Hconf\frac{\ddd\hat{\bar
E}_{ij}}{\ddd \eta} +k^2\hat{\bar
E}_{ij}=\Hconf_0^2 \hat{\bar\Pi}^s_{ij}.
\end{equation}
where a barred tensor is traceless and orthogonal to $k^i$.

The background quantity $\Hconf(\eta)$ being known from
\EQNS{\ref{h2}, \ref{eta}} and (\ref{pert_nu}-\ref{pert_tens}) give
the evolution of the perturbations in function of $\eta$ and the
comoving wavenumber $k$, once the source variables (constrained by
\EQNS{\ref{def_cons_s0}-\ref{def_cons_vi}}), are known in function of
$\eta$ and $k$ (\S \ref{sec_stress}) and once the initial conditions are
set (\S \ref{sec_ic}).

\section{The stress-energy tensor $\Theta_{\mu\nu}$ of the sources}
\label{sec_stress}

\subsection{Homogeneity and isotropy constraints}

The ten components of the stress-energy tensor
$\Theta_{\mu\nu}(\eta,x^i)$ of the topological defects
($\mu,\nu=0,1,2,3$) are ten statistically spatially homogeneous and
isotropic random fields. So are their ten Fourier transforms
$\hat\Theta_{\mu\nu}(\eta,k^i)$ (which are complex but such that
$\hat\Theta^*_{\mu\nu}(\eta,k^i)=\hat\Theta_{\mu\nu}(\eta,-k^i)$). (We
ignore the $k^i=0$ mode which can be absorbed in the background.)

The statistical properties of those ten random fields will be
described by their unequal time two-point correlators
\begin{equation}
\langle \Theta_{\mu\nu}(\eta, x^i)
\Theta_{\rho\sigma}(\eta',x'^i)\rangle \equiv
\frac{{\mathcal H}_0}{\kappa^2}
C_{\mu\nu\rho\sigma}(\eta,\eta',r^i),
\end{equation}
where $\langle ...\rangle $ means an ensemble average on a large
number of realisations, and where the correlator
$C_{\mu\nu\rho\sigma}$ is a tensor which depends only on $\eta, \eta'$
and $r^i\equiv x^i- x'^i$ because of the spatial homogeneity of the
distribution.  The power spectra of the correlators
$C_{\mu\nu\rho\sigma}$ are defined as
\begin{equation}
P_{\mu\nu\rho\sigma}(\eta,\eta',k^i)\equiv (2\pi)^{3/2}\hat
C_{\mu\nu\rho\sigma}(\eta,\eta',k^i),
\end{equation}
where a hat denotes a Fourier transform. The power spectra are related
to the correlators in Fourier space by
\begin{equation}
\label{corr_theta}
\langle \hat \Theta^*_{\mu\nu}(\eta,k^i)\hat
\Theta_{\rho\sigma}(\eta',k'^i)\rangle =
\delta ( k^i- k'^i) \frac{{\mathcal H}_0^4}{\kappa^2}
P_{\mu\nu\rho\sigma}(\eta,\eta',k^i).
\end{equation}
The spatial isotropy of the distribution now forces the power spectra
to be of the form 
\begin{eqnarray}
\label{p0000}
P_{0000} & = & A_0, \\
P_{000i} & = & ik_iB_1,\\
P_{00ij} & = & C_0\delta_{ij}+C_2k_ik_j, \\
P_{0i0j} & = & D_0\delta_{ij}+D_2k_ik_j, \\
P_{0ijk} & = & i\left[E_1k_i\delta_{jk} +\bar E_1(k_j\delta_{ik}
+k_k\delta_{ij})+E_3k_ik_jk_k\right], \\
\label{pijkl}
P_{ijkl} & = & F_0\delta_{ij}\delta_{kl}+\bar
F_0(\delta_{ik}\delta_{jl}+\delta_{il}\delta_{jk})+
F_2(k_ik_j\delta_{kl}+k_kk_l\delta_{ij})+ \\
\nonumber
 & & \bar F_2
(k_ik_k\delta_{jl}+k_ik_l\delta_{jk}+k_jk_l\delta_{ik}+k_jk_k\delta_{il})+
F_4k_ik_jk_kk_l,
\end{eqnarray}
where $A_0, B_1$ \ETC{} are 14 real functions of $\eta$, $\eta'$ and
the modulus $k$ of the spatial vector $k^i$.

When $\Theta_{\mu\nu}$ is decomposed into its scalar, vector and
tensor components according to \EQ{\ref{svt_00}-\ref{svt_ij}}, then
\EQNS{\ref{corr_theta}} and (\ref{p0000}-\ref{pijkl}) yield
\begin{eqnarray}
\label{rhorho}
\langle\hat\rho^*_s\hat\rho_s\rangle & = & A_0, \\
\langle\hat\rho^*_s\hat V_s\rangle   & = & -k B_1, \\
\langle\hat\rho^*_s\hat P_s\rangle   & = & C_0+\frac{1}{3}k^2C_2, \\
\langle\hat\rho^*_s\hat\pi_s\rangle  & = & -k^2C_2, \\
\langle\hat V^*_s\hat V_s\rangle     & = & D_0+k^2D_2, \\
\langle\hat V^*_s\hat P_s\rangle     & = & 
  k\left[\left(E_1+\frac{2}{3}\bar E_1\right)+\frac{k^2}{3}E_3\right], \\
\langle\hat V^*_s\hat \pi_s\rangle   & = & -k(2\bar E_1+k^2E_3), \\
\langle\hat P^*_s\hat P_s\rangle     & = & 
  \left(F_0+\frac{2}{3}\bar F_0\right)+\frac{2}{3}k^2
  \left(F_2+\frac{2}{3}\bar F_2\right)+\frac{1}{9}k^4F_4, \\
\langle\hat P^*_s\hat \pi_s\rangle   & = & 
  -k^2\left[\left(F_2+\frac{4}{3}\bar
  F_2\right)+\frac{1}{3}k^2F_4\right], \\
\label{pispis}
\langle\hat\pi^*_s\hat\pi_s\rangle   & = & 3\bar F_0+4k^2\bar F_2+k^4F_4,
\end{eqnarray}
\begin{eqnarray}
\label{vvvv}
\langle \hat {\bar v}^{s*}_i\hat{\bar v}_j^s\rangle   & = & 
  D_0 P_{ij}, \\
\langle \hat {\bar v}^{s*}_i\hat{\bar \pi}_j^s\rangle & = & 
  -k\bar E_1P_{ij}, \\
\label{pivpiv}
\langle\hat{\bar\pi}^{s*}_i\hat{\bar\pi}_j^s\rangle   & = & 
  (\bar F_0 + k^2 \bar F_2) P_{ij},
\end{eqnarray}
\begin{eqnarray}
\label{pitpit}
\langle \hat {\bar \Pi}^{s*}_{ij} \hat {\bar \Pi}_{kl}^s\rangle & = & 
\bar F_0(P_{ik}P_{jl}+P_{il}P_{jk}-P_{ij}P_{kl}),
\end{eqnarray}
where $\langle\hat\rho^*_s(\eta,k^i)\hat\rho_s(\eta',k'^i)\rangle
=\delta(k^i-k'^i) \langle\hat\rho^*_s\hat\rho_s\rangle$ \ETC, and
where $P_{ij}\equiv \delta_{ij}-k_ik_j/k^2$.  All other correlators
are zero so that the scalar, vector and tensor parts of
$\Theta_{\mu\nu}$ fall into statistically independent sets.

We also note that {\it if} $(k^2 \bar F_2)$ and $(k^4 F_4)$ are of
higher order in $k$ than $\bar F_0$, then, for small $k$
\cite{pst1}\cite{turok}
\begin{equation}
\label{pipi324}
\frac{1}{3}\langle\hat\pi^*_s\hat\pi_s\rangle
\simeq\frac{1}{2} \langle \hat {\bar \pi}^{s*}_i\hat 
{\bar\pi}^i_s\rangle \simeq \frac{1}{4}\langle \hat {\bar
\Pi}^{s*}_{ij}\hat {\bar\Pi}^{ij}_s\rangle.
\end{equation}

\subsection{Causality constraints}

Since the network of defects appeared at a definite time the
distribution must be, for causality reasons, completely uncorrelated
on scales larger than the particle horizon.  Therefore, as stressed
\EG{} by Turok \cite{turok}, the unequal time correlators are strictly
zero outside the intersection of the past light-cones, that is~:
\begin{equation}
\label{corr_causal}
C_{\mu\nu\rho\sigma}(\eta,\eta',r^i) =0
\qquad\hbox{if}\qquad r>\eta+\eta'.
\end{equation}

Property (\ref{corr_causal}) translates in Fourier space into the fact
that the equal time power spectra are white noise on super horizon
scales (that is for $k\eta\ll1$). Indeed, because the correlators
(\ref{corr_causal}) have compact supports their Fourier transforms are
$C^\infty$ in $k^i$. Therefore causality forces the fourteen functions
$A_0$, $B_1$ \ETC{} to be $C^\infty$ in $k^2$.

Moreover, since within one horizon volume there are almost no defects,
those fourteen functions must tend to zero on small scales, that is
for $k\eta\gg1$.

\subsection{Coherence hypothesis}
\label{ssec_coh}

Any distribution of active sources such as topological defects must be
homogeneous, isotropic and causal. The hypothesis of statistical
coherence, that we shall now make, is stringent but motivated by the
fact that any incoherent distribution can in principle be decomposed
into a sum of coherent ones \cite{pst1}.

By definition two statistically homogeneous random fields $\hat
S_1(\eta,k^i)$ and $\hat S_2(\eta, k^i)$ are statistically coherent if
their correlators factorize, that is if 

\begin{equation}
\label{corr_coh}
\langle\hat S^*_a(\eta,k^i)\hat
S_b(\eta',k'^i)\rangle=\delta(k^i-k'^i) p_a(\eta,k^i)
p_b(\eta',k^i)
\end{equation}
for $a,b=1,2$. For our purposes this is equivalent to saying that
\begin{equation}
\label{corr_coh2}
\hat S_a(\eta,k^i)=p_a(\eta, k^i)e(k^i)
\end{equation}
where $p_a(\eta, k^i)$ is a real function and where $e(k^i)$ is a
normalized complex random field such that $\langle
e^*(k^i)e(k'^i)\rangle=\delta(k^i-k'^i)$.

Let us first combine coherence and isotropy, that is
(\ref{corr_coh}-\ref{corr_coh2}) and (\ref{rhorho}-\ref{pitpit}). The four
scalar components $\hat\rho^s$, $\hat V^s$, $\hat P^s$, and
$\hat\pi^s$ of $\hat\Theta_{\mu\nu}$ form a statistically independent
set. The hypothesis (\ref{corr_coh2}) of maximal coherence implies that
\begin{eqnarray}
\hat\rho^s(\eta,k^i)& = & p_\rho(\eta,k) e(k^i), \\
\hat V^s(\eta,k^i) & = & p_V(\eta,k)e(k^i), \\
\hat P^s(\eta,k^i) & = & p_P(\eta,k)e(k^i), \\
\hat\pi^s(\eta,k^i) & = & p_\pi(\eta,k)e(k^i).
\end{eqnarray}
Hence, out of the fourteen functions $A_0$, $B_1$ \ETC{} appearing in
(\ref{rhorho}-\ref{pispis}) ten can be expressed in terms of the four
real functions $p_a$ and the four functions (\EG{} $D_0$, $\bar E_1$,
$\bar F_0$ and $\bar F_2$) that remain arbitrary. More precisely~:
\begin{eqnarray}
\label{corr_coh_a0}
A_0 & = & p_\rho^2, \\
B_1 & = & -p_\rho \left({p_V\over k}\right), \\
C_0 & = & p_\rho\left(p_P+{1\over3}p_\pi\right), \\
C_2 & = & -p_\rho\left({p_\pi\over k^2}\right), \\
E_1 & = & {p_V\over k}\left(p_P+{1\over3}p_\pi\right), \\
\label{corr_coh_d2}
D_2 & = & \left({p_V\over k}\right)^2-{D_0\over k^2}, \\ E_3 & = &
-\left({p_V\over k}\right)\left({p_\pi\over k^2}\right) -2{\bar
E_1\over k^2}, \\ F_2 & = & -{p_\pi\over
k^2}\left(p_P+{1\over3}p_\pi\right) +{\bar F_0\over k^2}, \\ 
F_0 & = & \left(p_P+{1\over3}p_\pi\right)^2-\bar F_0, \\
\label{corr_coh_f4}
F_4 & = & 
  \left({p_\pi\over k^2}\right)^2-3{\bar F_0\over k^4}-4{\bar F_2\over k^2}.
\end{eqnarray}

The four vector components $\hat{\bar v}_i^s$ and $\hat{\bar\pi}^s_i$
of $\hat\Theta_{\mu\nu}$ form yet another statistically independent
set. In order to obey (\ref{corr_coh2}) as well as
(\ref{vvvv}-\ref{pivpiv}) they must fall into two independent subsets
corresponding to two different polarisations. Hence we shall write
them as
\begin{eqnarray}
\label{corr_coh_v}
\hat{\bar v}_i^s(\eta,k^i)  & = & p_{\bar v}(\eta,k)\bar e_i(k^i), \\
\hat{\bar\pi}^s_i(\eta,k^i) & = & p_{\bar \pi}(\eta,k)\bar e_i(k^i), \\
\bar e_i(k^i) & \equiv & \bar l_ie_l(k^i)+\bar m_ie_m(k^i),
\end{eqnarray}
where $\langle e^*_A(k'^i)e_B(k^i)\rangle=\delta_{AB}\delta(k'^i-k^i)$
with $A,B$ standing for $l$ or $m$, and where $\bar l^i$ and $\bar
m^i$ form with $k^i/k$ an orthonormal basis (so that $P_{ij}=\bar l_i
\bar l_j + \bar m_i\bar m_j$).  Hence two of the three previously free
functions $D_0$, $\bar E_1$ and $\bar F_0$ can be expressed in terms
of the two real functions $p_{\bar V}$ and $p_{\bar \Pi}$ as [see
(\ref{vvvv}-\ref{pivpiv})]
\begin{eqnarray}
\label{corr_coh_d0}
D_0      & = & p_{\bar v}^2, \\
\bar E_1 & = & -{p_{\bar v}\over k}p_{\bar \pi}, \\
\bar F_0 & = & p_{\bar \pi}^2 - k^2 \bar F_2.
\end{eqnarray} 

Finally the two tensor components $\hat{\bar\Pi}^s_{ij}$ of
$\hat\Theta_{\mu\nu}$ form a third statistically independent set. In
order to obey (\ref{pitpit}) they cannot be coherent but, on the
contrary, independent of each other. Hence we shall write them as
\begin{eqnarray}
\label{corr_coh_t}
\hat{\bar\Pi}^s_{ij}(\eta,k^i) & =      & p_T(\eta,k)\bar e_{ij}(k^i), \\
e_{ij}(k^i) & \equiv & \bar l_{ij}e_+(k^i)+  \bar m_{ij}e_\times(k^i),
\end{eqnarray}
where $\bar l_{ij}\equiv\bar l_i\bar l_j-\bar m_i\bar m_j$ and $\bar
m_{ij}\equiv\bar l_i\bar m_j+\bar l_j\bar m_i$ are the two
polarisation tensors, where $\langle
e^*_A(k'^i)e_B(k^i)\rangle=\delta_{AB}\delta(k'^i-k^i)$ with $A,B$
standing for $+$ or $\times$, and where [see \EQNS{\ref{pitpit}}]
\begin{equation}
\label{corr_coh_fb0}
\bar F_0 = p_T^2.
\end{equation}

In conclusion combining coherence and isotropy reduces the number of
free functions from fourteen to seven~: $p_\rho,p_V,p_P,p_\pi,p_{\bar
v}$, $p_{\bar\pi}$ and $p_T$.

Let us add now the causality constraint which states that the fourteen
functions $A_0$, $B_1$ \ETC{} must be $C^\infty$ in $k^2$. It implies
[see (\ref{corr_coh_a0})] that $p_\rho$ , $p_V/ k$ , $p_P$ and $p_\pi/
k^2$ are $C^\infty$. This in turn implies [see
\EQNS{\ref{corr_coh_d2}-\ref{corr_coh_f4}}] that $D_0/k^2$ , $\bar
E_1/k^2$ , $\bar F_2/k^2$ and $\bar F_0/k^4$ are analytic as well, and
therefore, from (\ref{corr_coh_d0}), $p_{\bar v}/k$ and
$p_{\bar\pi}/k^2$ are $C^\infty$. Finally, from (\ref{corr_coh_fb0}),
$p_T$ is also $C^\infty$.

An immediate consequence of this regularity requirement is that
\begin{eqnarray}
\langle\hat\pi^*_s\hat\pi_s\rangle & = & p_\pi^2={\mathcal O}(k^{4}), \\
\langle \hat {\bar\pi}^{s*}_i\hat {\bar \pi}^i_s\rangle & = &
  2p_{\bar\pi}^2={\mathcal O}(k^4), \\
\langle \hat {\bar\Pi}^{s*}_{ij}\hat  {\bar\Pi}^{ij}_s\rangle & = & 
  4p_T^2={\mathcal O}(k^4),
\end{eqnarray}
so that, contrarily to the general case (\ref{pipi324}), the
anisotropic stress correlators of coherent defects (when $p_\pi\neq
p_{\bar\pi} \neq p_T$) are not in a definite ratio for small $k$.

A question that can be asked is how can a sum of coherent defects
(with anisotropic stress correlators of order $k^4$) leads to an
incoherent defect (with anisotropic stress correlator of order $k^0$).
In a paper to be submitted \cite{rdxx}, we show explicitely how the
$k^0$-behaviour of incoherent defects can be obtained as a (infinite)
sum of $k^4$-behaving coherent defects. We also show how the causality
constraints fade away in the process and how the procedure introduces
almost no extra arbitrariness.

\subsection{Scaling hypothesis}

We shall now make the other assumption that, as long as the evolution
of the universe is described by a single scale (to wit the Hubble
radius ${\mathcal H}^{-1}$), that is deep in the radiation era and
deep in the matter era but not during the transition, the
dimensionless quantities $\Theta_\mu^\nu/\rho_t$ are scale invariant,
that is such that
\begin{equation}
C_{\mu\nu\mu\nu}(\eta,\eta,r^i)={\mathcal H}^4F_{\mu\nu}(r^i{\mathcal H})
\end{equation}
(using the fact that $\kappa a^2\rho_t\propto {\mathcal H}^2$ in both
eras).  This assumption is supported by a number of numerical
simulations \cite{def_num1}-\cite{def_num5}\cite{durrer}\cite{pspt}
as well as qualitative arguments \cite{durrer}\cite{bray}.  In
Fourier space this hypothesis translates as
\begin{equation}
P_{\mu\nu\mu\nu}(\eta,\eta, k^i)=(2\pi)^{3/2}{\mathcal H}\hat
F_{\mu\nu}(k^i/{\mathcal H}).
\end{equation}
This implies [see (\ref{p0000}-\ref{vvvv})] that the functions
$A_0/{\mathcal H}$, $D_0/{\mathcal H}$, $F_0/{\mathcal H}$, $\bar
F_0/{\mathcal H}$ on one hand, $D_2{\mathcal H}$, $F_2{\mathcal H}$,
$\bar F_2{\mathcal H}$ on another and finally $F_4{\mathcal H}^3$ are
all functions of $k/{\mathcal H}$ only (when $\eta=\eta'$). From
\EQNS{\ref{rhorho}-\ref{pitpit}} one then obtains the behaviour of
the equal time correlators $\langle\hat\rho^*_s\hat\rho_s\rangle$,
$\langle\hat V^*_s\hat V_s\rangle$, $\langle\hat P^*_s\hat
P_s\rangle$, $
\langle\hat P^*_s\hat \pi_s\rangle$, $
\langle\hat\pi^*_s\hat\pi_s\rangle$, $
\langle \hat {\bar v}^{s*}_i\hat{\bar v}_j^s\rangle$, $
\langle \hat {\bar \pi}^{s*}_i\hat{\bar \pi}_j^s\rangle$ 
and $\langle \hat {\bar\Pi}^{s*}_{ij}\hat {\bar \Pi}_{kl}^s\rangle$.

Combining this scaling hypothesis with the one of coherence implies
[see \EQNS{\ref{corr_coh_a0})-(\ref{corr_coh_f4}}] that all six
$p_a/\sqrt{\mathcal H}$ are functions of $k/{\mathcal H}$ only.

Therefore, all in all, coherent defects, which scale during the
radiation and matter dominated eras, which are causal and
statistically homogeneous and isotropic, are described by the
following random variables~:
\begin{eqnarray}
\label{corr_coh_rho}
\hat\rho^s           & = & {\sqrt h}\,f_1(q/h, x)\,e(q^i), \\
\hat P^s             & = & {\sqrt h}\,f_2(q/h, x)\,e(q^i), \\
\hat V^s             & = & -{q\over {\sqrt h}}\,f_3(q/h, x)\,e(q^i), \\
\hat \pi^s           & = & {q^2\over h^{3/2}}\,f_4(q/h, x)\,e(q^i), \\
\label{corr_coh_vv}
\hat{\bar v}_i^s     & = & {q\over {\sqrt h}}\,f_5(q/h, x)\,\bar e_i(q^i), \\
\hat{\bar\pi}^s_i    & = & {q^2\over h^{3/2}}\,f_6(q/h, x)\,\bar e_i(q^i), \\
\label{corr_coh_pit}
\hat{\bar\Pi}^s_{ij} & = & {q^2\over h^{3/2}}\,f_7(q/h, x)\,\bar e_{ij}(q^i)
\end{eqnarray}
where we have reintroduced the dimensionless quantities $x\equiv
a/a_0$, $h\equiv{\mathcal H}/{\mathcal H}_0$ and $q^i\equiv
k^i/{\mathcal H}_0$, and where the seven dimensionless functions
$f_a(q/h, x)$ depend only on $q/h$ deep in the radiation or matter
eras, tend to constants on super-horizon scale (that is for
$k\eta\propto q/h\ll1$) and to zero on small scales.

\subsection{Conservation laws and loss of scaling invariance 
during the radiation to matter dominated transition}

When the defects are described by
\EQNS{\ref{corr_coh_rho}-\ref{corr_coh_pit}} the conservation
equations (\ref{def_cons_s0}-\ref{def_cons_vi}) become
\begin{eqnarray}
\label{def_cons_scal_s0}
{1\over2}(1+3\omega)u{\partial f_3\over\partial u}
+x{\partial f_3\over \partial x}
+{3\over4}(3+\omega)f_3
 & = & f_2-{2\over3}u^2f_4, \\
{1\over2}(1+3\omega)u{\partial f_1\over\partial u}
+x{\partial f_1\over\partial x} 
+{3\over4}(1-\omega)f_1
 & = & -3f_2-u^2f_3, \\
\label{def_cons_scal_vi}
{1\over2}(1+3\omega)u{\partial f_5\over\partial u}
+x{\partial f_5\over\partial x}
+{3\over4}(3+\omega)f_5
 & = & u^2f_6,
\end{eqnarray}
where we have introduced
\begin{eqnarray}
u      & \equiv & q/h, \\
\label{omega}
\omega & \equiv & {1\over3}{\Omega_r^0\over\Omega_r^0+x\Omega_m^0}.
\end{eqnarray}

Deep in the radiation or matter eras $\omega\to 1/3$ or $0$, and
scaling solutions independent of $x$ are readily found for $f_2$,
$f_4$ and $f_6$ in terms of the four free functions $f_1$, $f_3$ and
$f_5$. During the transition from the radiation to matter dominated
eras, scaling is lost. How can this loss of scaling be modelled, and
how is the final result dependent on this modelisation~?

Two different approaches have been followed~:
\begin{enumerate}
  \item \label{ap1} Durrer \ETAL{} \cite{durrer} used a ``sudden
  transition'' approximation in which $\omega=1/3$ until $x=x_{\rm
  eq}$ and $\omega=0$ afterwards. The problem here is that the
  functions $f_a$ cannot all be continuous at $x=x_{\rm eq}$ and that
  energy momentum conservation is violated during the
  transition\footnote{Note added in proof~: In Durrer \ETAL{}
  \cite{dkm}, the loss of scaling is taken into account by
  interpolating the source functions computed respectively in the
  radiation and matter era. The fit is performed by using either the
  function $t_{\rm eq}/(t+t_{\rm eq})$ or $exp(-t/t_{\rm eq})$ and the
  results do not depend too much on this interpolation. However,
  energy conservation is still violated during the transition.}.

  \item \label{ap2} Turok \cite{turok}, Hu \ETAL{} \cite{huspwh},
  Cheung and Magueijo \cite{cheungma} fix the time dependence of two
  source functions and determine the other two by integrating
  (\ref{def_cons_scal_s0}-\ref{def_cons_scal_vi}).
\end{enumerate}
We follow a route similar to (\ref{ap2}), which also amounts to
making fairly artificial ans\"atze but which garantees that the
conservation equations are exactly satisfied~: we first impose
\begin{eqnarray}
f_1 & = & {\mathcal F}(u), \\
f_3 & = & \tilde f_3(u)g(x), \\
f_5 & = & \tilde f_5(u)p(x), \\
f_7 & = & {\mathcal Q}(u).
\end{eqnarray}
It then follows from (\ref{def_cons_scal_s0}-\ref{def_cons_scal_vi})
that
\begin{eqnarray}
\tilde f_3(u) & = & -{(1-\omega){\mathcal F}\over3(3+\omega)g+4x\ddd g/\ddd x}
                    +u^2{\mathcal G}(u,x) \\
\tilde f_5(u) & = & \frac{u^2{\mathcal P}(u,x)}{3(3+\omega)p+4x\ddd p/\ddd x}
\end{eqnarray}
where ${\mathcal G}(u,x)$ and ${\mathcal P}(u,x)$ are some functions,
$C^\infty$ in $u^2$, that we impose to depend on $u$ only
\begin{equation}
{\mathcal G}(u,x)={\mathcal G}(u)\quad,\quad {\mathcal
P}(u,x)={\mathcal P}(u).
\end{equation}

This implies that $g(x)$ and $p(x)$ satisfy the following differential
equations
\begin{equation}
\label{jpu_s_g}
(3+\omega)g+{4\over3}x{\ddd g\over \ddd x}=1-\omega
\end{equation}
\begin{equation}
\label{jpu_v_p}
(3+\omega)p+{4\over3}x{\ddd p\over \ddd x}=1
\end{equation}
(up to irrelevant overall constants) with $\omega$ given by
(\ref{omega}) and $g(x\to0)=1/5$, $p(x\to0)=3/10$ (and
$g(x\to\infty)=p(x\to\infty)=1/3$).  We hence arrive at the following
expressions for the seven functions $f_a$ in terms of the four
arbitrary $C^\infty$ functions ${\mathcal F}(u)$, ${\mathcal G}(u)$,
${\mathcal P}(u)$ and ${\mathcal Q}(u)$, with $g(x)$ and $p(x)$ being
given by (\ref{jpu_s_g}-\ref{jpu_v_p})~:
\begin{eqnarray}
\label{jpu_f1}
f_1 & = & {\mathcal F}, \\
f_2 & = & -{1\over4}(1-\omega){\mathcal F}
          -{1\over6}(1+3\omega)u{d{\mathcal F}\over du}
           +{u^2\over9}g{\mathcal F}-{u^4\over3}g{\mathcal G}, \\
f_3 & = & \left[-{1\over3}{\mathcal F}+u^2{\mathcal G}\right]g, \\
\label{jpu_f4}
f_4 & = & {1\over4}(1+3\omega)
           \left[(g-1){1\over u}{d{\mathcal F}\over du}
                 -6{\mathcal G}g-3gu{d{\mathcal G}\over du}\right]\\
\nonumber    &   & +{1\over6}g{\mathcal F}-{9\over8}(1-\omega){\mathcal G}
          -{u^2\over2}{\mathcal G}g, \\
f_5 & = & {1\over3}u^2{\mathcal P} p, \\
f_6 & = & \left[{1\over4}+(1+3\omega){p\over3}\right]{\mathcal P}
          +{1\over6}(1+3\omega)pu{d{\mathcal P}\over du}, \\
\label{jpu_f7}
f_7 & = & {\mathcal Q}.
\end{eqnarray}

In conclusion, we note that in our model both components of the
defects stress-energy tensor are generically present, in contrary to
\EG{} the ``pressure model'' or the ``anisotropic stress model''
\cite{huspwh}\cite{cheungma}.

\section{Initial conditions}
\label{sec_ic}

The set of equations (\ref{pert_nu}-\ref{pert_tens}) with the defects
described by \EQNS{\ref{corr_coh_rho}-\ref{corr_coh_pit}} and
(\ref{jpu_f1}-\ref{jpu_f7}) will be solved numerically. However, deep
in the radiation era when all scales of cosmological interest today
were larger than the Hubble radius, an analytic solution exists, which
is easily found when adding to the set (\ref{pert_nu}-\ref{pert_tens}) the
Einstein equations for scalar perturbations which we had not included
because they were redundant (by virtue of the Bianchi
identities). Those two extra equations can be written as~:
\begin{eqnarray}
\label{poisson_der_1}
x{\ddd\hat\Psi\over \ddd x} & = & -\hat\Phi-{\hat V^s\over2qh} \\
\nonumber&&
-{1\over2qhx^2}[4(\Omega_\nu^0\hat V_\nu+\Omega_\gamma^0\hat V_\gamma)
                +3x(\Omega_b^0\hat V_b+\Omega_c^0\hat V_c)],\\
\label{poisson_der_2}
(\Omega_r^0+x\Omega_m^0)\frac{\ddd^2\hat\Psi}{\ddd x^2} 
 & = & -{1\over x}{\ddd\hat\Psi\over \ddd x} 
       \left(4\Omega_r^0+ {9\over2}x\Omega_m^0\right) \\
\nonumber&&   -   {\hat\Psi\over x^2}
       \left({5\over2}\Omega_m^0x+{1\over3}q^2x^2\right) \\
\nonumber&   &
+{1\over q^2x}\left[{\ddd\hat\pi^s\over \ddd x}(\Omega_r^0+x\Omega_m^0)
                   +\Omega_m^0\hat\pi^s\right]\\
\nonumber&&
-{1\over2x}(\Omega_b^0 \hat\delta^\flat_b+\Omega_c^0\hat\delta^\flat_c) \\
\nonumber& &
 -\left({\hat\pi^s\over3}+{\hat\rho^s\over6}-{\hat P^s\over2}\right).
\end{eqnarray}

Deep in the radiation era, when $x\Omega_m^0\ll\Omega_r^0$~:
$hx\simeq\sqrt{\Omega_r^0}$ (and $x\propto\eta$). The wavelengths we
are interested in are then much bigger than the Hubble radius~:
$q/h\ll1$, that is $q^2x^2\ll\Omega_r^0$ (or $u^2\ll1$). In that
regime the conservation equations for the defects
[\EQNS{\ref{jpu_f1}-\ref{jpu_f7}}] yield that the functions $f_a$
are constants such that
\begin{eqnarray}
f_1 & = & -6f_2+{\mathcal O}(u^2), \\
f_3 & = & {2\over5}f_2+{\mathcal O}(u^2), \\
f_5 & = & {2\over9}u^2f_6+{\mathcal O}(u^4).
\end{eqnarray}

Let us first solve the equations for the scalar perturbations. The
conservation equations for the fluids (\ref{pert_nu}, \ref{pert_cdm},
\ref{pert_bar_avd_scal}-\ref{pert_pb_avd_vec}) impose that in the long
wavelength limit all the density perturbations $\hat\delta^\flat_n$
are constant, that is random variables independent of time.
\begin{eqnarray}
\label{ci_delta_nu}
\hat\delta^\flat_\nu    & \approx & {\delta^\flat_\nu}^{\rm in}, \\
\hat\delta^\flat_c      & \approx & {\delta^\flat_c}^{\rm in}, \\
\hat\delta^\flat_b      & \approx & {\delta^\flat_b}^{\rm in}, \\
\hat\delta^\flat_\gamma & \approx & {\delta^\flat_\gamma}^{\rm in}
\end{eqnarray}
where $\approx$ means deep in the radiation era and up to terms of
order $u^2$. \EEQ{\ref{poisson_der_2}} for $\hat\Psi$ is therefore a
closed equation in that regime. Its solution is
\begin{eqnarray}
\label{ci_psi}
\hat\Psi & \approx & 
\left[\Psi_0+{\Psi_1\over x^3}
      +{2\over9}x^{3/2}(\tilde f_4+\tilde f_2)e(q^i)
      -{{\delta^\flat_m}^{\rm in}\over 8\Omega_r^0}x\right], \\
{\delta^\flat_m}^{\rm in} & \equiv & 
  \Omega_b^0{\delta^\flat_b}^{\rm in}+\Omega_c^0\delta^\flat_{c{\rm
in}}
\end{eqnarray}
where $\Psi_0$ and $\Psi_1$ are two integration constant random
variables and where we have set $\tilde
f_a\equiv{\Omega_r^0}^{-3/4}\lim_{u\to0}f_a$. The perturbation
$\hat\Phi$ then follows from \EQ{\ref{poisson2}}
\begin{equation}
\hat\Phi\approx\left[\Psi_0+{\Psi_1\over
x^3}+{1\over9}x^{3/2}(2\tilde f_2-7\tilde f_4)e(q^i)-{
{\delta^\flat_m}^{\rm in}\over 8\Omega_r^0}x\right].
\end{equation}

The Euler equations (\ref{pert_nu}, \ref{pert_cdm},
\ref{pert_bar_avd_scal}-\ref{pert_pb_avd_vec}) then give the velocity
perturbations of the fluids
\begin{eqnarray}
{\hat V_\nu\over q} & \approx & { V_\nu^{\rm in}\over
q}-\left.{1\over\sqrt{\Omega_r^0}}
\right[\left({1\over4}{\delta^\flat_\nu}^{\rm in}+2\Psi_0\right)x
-{\Psi_1\over x^2} \\
\nonumber & & 
\left.+{2\over45}x^{5/2}(4\tilde f_2-5\tilde f_4)e(q^i)
-{{\delta^\flat_m}^{\rm in}\over 8\Omega_r^0}x^2\right], \\
{\hat V_c\over q} & \approx & 
{ V_c^{\rm in}\over xq}-\left.{1\over\sqrt{\Omega_r^0}}
  \right[{\Psi_0\over2}x-{\Psi_1\over x^2} \\
\nonumber& &\left.
        +{2\over63}x^{5/2}(2\tilde f_2-7\tilde f_4)e(q^i)
        -{{\delta^\flat_m}^{\rm in}\over 24\Omega_r^0}x^2\right], \\
\label{ci_vs_bar}
{\hat V_b\over q}={\hat V_\gamma\over q} & \approx & 
{ V_\gamma^{\rm in}\over q}- \left.{1\over\sqrt{\Omega_r^0}}
\right[\left({1\over4}{\delta^\flat_\gamma}^{\rm in}
      +2\Psi_0\right)x-{\Psi_1\over x^2} \\
\nonumber&&\left.
      +{2\over45}x^{5/2}(4\tilde f_2-5\tilde f_4)e(q^i)
      -{{\delta^\flat_m}^{\rm in}\over 8\Omega_r^0}x^2\right]
\end{eqnarray}
where $ V_\nu^{\rm in}$, $ V_c^{\rm in}$ and $ V_\gamma^{\rm
in}$ are constants of integration. The nine integration constants
introduced (the four density perturbations ${\delta^\flat_n}^{\rm
in}$, $\Psi_0$, $\Psi_1$ and the three initial velocity perturbations
$V_n^{\rm in}$) are constrained by the remaining equations
(\ref{poisson_der_1}) and (\ref{poisson})
\begin{eqnarray}
\Psi_0 & = & 
-{1\over6\Omega_r^0}(\Omega_\gamma^0{\delta^\flat_\gamma}^{\rm in}
+\Omega_\nu^0{\delta^\flat_\nu}^{\rm in}), \\
V_\nu^{\rm in} & = & - V_\gamma^{\rm in},
\end{eqnarray}
so that the system depends on seven integration constants as
requested.

Solving the equations for the vector perturbations, deep in the
radiation era and for long wavelengths, is straightforward. The
conservations equations (\ref{pert_nu}, \ref{pert_cdm},
\ref{pert_bar_avd_scal}-\ref{pert_pb_avd_vec}) yield
\begin{eqnarray}
\label{ci_vv_nu}
\hat{\bar v}^i_{\nu} & = & \hat{\bar v}_\nu^{i\,\rm in}, \\
\hat{\bar v}^i_{c}   & = & {\hat{\bar v}_c^{i\,{\rm in}}\over x}, \\
\label{ci_vv_pb}
\hat{\bar v}^i_{b} = \hat{\bar v}^i_{\gamma} 
 & = & \hat{\bar v}_\gamma^{i\, {\rm in}}
\end{eqnarray}
where $\hat{\bar v}_n^{i\,{\rm in}}$ are three constant vectors
orthogonal to $k^i$, and \EQ{\ref{poisson_vec}} gives
\begin{equation}
\label{ci_phi_v}
\hat{\bar\Phi}^i\approx\left[{\hat{\bar\Phi}^i_0\over
x^2}+{\hat{\bar\Phi}^i_1\over
x}+{4\over9}{q\over\sqrt{\Omega_r^0}}x^{5/2}\tilde f_6\bar
e^i(q^i)\right]
\end{equation}
with $\hat{\bar\Phi}^i_0\equiv(2/q^2)[3\Omega_c^0\hat{\bar
v}_c^{i\,{\rm in}}+4(\Omega_\gamma^0 \hat{\bar v}^i_{\gamma {\rm
in}}+\Omega_\nu^0\hat{\bar v}_\nu^{i\,{\rm in}})]$, and
$\hat{\bar\Phi}^i_1\equiv 6\Omega_b^0\hat{\bar v}_\gamma^{i\,{\rm in}}
/q^2$.

Finally the tensor perturbations solve \EQ{\ref{pert_tens}} [with
$\bar\Pi^s_{ij}$ given by (\ref{corr_coh_pit})]~:
\begin{equation}
\label{ci_eij}
\bar E^{ij}\approx\left[\bar E^{ij}_0+{\bar E^{ij}_1\over x}+{4\over
63}{q^2\over\Omega_r^0}x^{7/2}\tilde f_7\bar
e^{ij}(q^i)\right]
\end{equation}
where $\bar E^{ij}_0$ and $\bar E^{ij}_1$ are two constant traceless
tensors orthogonal to $k^i$.
 
The solution we have obtained is valid deep in the radiation era, for
wavelengths larger than the Hubble radius, but late enough so that the
baryons and CDM are non-relativistic. At the epoch of the phase
transition which gave rise to the defects all the matter was
relativistic and the solution (given in \cite{dlu}) is
(\ref{ci_delta_nu}-\ref{ci_eij}) where one sets
$\Omega_b^0=\Omega_c^0=\Omega_\nu^0=0$
and $\Omega_\gamma^0\equiv\Omega_r^0=1$.  As shown in \cite{dlu},
the simple fact that the defects suddenly appear in an up to then
perfectly homogeneous and isotropic universe in such a way that the
standard general relativistic matching conditions are satisfied
determines some of the constants of integration in terms of the
$\tilde f_a$. More precisely, two of the seven integration constants
which appear in the scalar perturbation set, one of the three
integration constant vectors for the vector set and the two
integration constant tensors for the tensor set are determined by
these matching conditions (see \cite{dlu} for details). The other
constants must be fixed by the physics at the epochs when the baryons
and CDM become non-relativistic. For example, if one assumes that the
transition is sudden and occurs on a hypersurface of constant density
then (see\cite{dlu}) $\Psi$ must be continuous and therefore
$\delta^\flat_{m{\rm in}}=0$.

It turns out however that almost all those initial conditions which
fix the integration constants are ``forgotten'' at late times, that is
long enough after the transitions. This is clearly seen for all the
metric perturbations and the scalar velocity perturbations. Indeed,
asymptotically [see \EQNS{\ref{ci_psi}-\ref{ci_vs_bar},
\ref{ci_phi_v}-\ref{ci_eij}}].
\begin{eqnarray}
\label{ci_res_deb}
\hat\Psi         & \to & {2\over9}x^{3/2}(\tilde f_4+\tilde f_2)e(q^i), \\
\hat\Phi         & \to & {1\over9}x^{3/2}(2\tilde f_2-7\tilde f_4)e(q^i), \\
\hat{\bar\Phi}_i & \to & {4\over9}{q\over\sqrt{\Omega_r^0}}x^{5/2}
                         \bar f_6\bar e_i(q^i), \\
\bar E_{ij}      & \to & {4\over 63}{q^2\over\Omega_r^0}x^{7/2}
                         \tilde f_7\bar e_{ij}(q^i), \\
\label{ci_res_v0}
\hat V_\nu \to \hat V_b=\hat V_\gamma & \to & 
  -                      {2\over45}{q\over\sqrt{\Omega_r^0}}x^{5/2}
                         (4\tilde f_2-5\tilde f_4)e(q^i), \\
\label{ci_res_v1}
\hat V_c         & \to & -{2\over63}{q\over\sqrt{\Omega_r^0}}x^{5/2}
                         (2\tilde f_2-7\tilde f_4)e(q^i).
\end{eqnarray}

In order now to obtain the asymptotic behaviours of the scalar density
perturbations one must solve the equations at next order in $q$. This
is easily done~: injecting the asymptotic expressions
(\ref{ci_res_v0}-\ref{ci_res_v1}) for the scalar velocity
perturbations into the conservation equations (\ref{pert_nu},
\ref{pert_cdm},
\ref{pert_bar_avd_scal}-\ref{pert_pb_avd_vec}) one gets
\begin{eqnarray}
    \hat\delta^\flat_\nu
\to \hat\delta^\flat_\gamma
\to {4\over3}\hat\delta^\flat_b 
 & \to & -{16\over945}{q^2\over\Omega_r^0}
         (4\tilde f_2-5\tilde f_4)x^{7/2}e(q^i), \\
\hat\delta^\flat_c & \to & -{4\over441}{q^2\over\Omega_r^0}
                           (2\tilde f_2-7\tilde f_4)x^{7/2}e(q^i).
\end{eqnarray}

Finally the solutions (\ref{ci_vv_nu}-\ref{ci_vv_pb}) hold for all
wavelengths so that the initial conditions for the vector velocity
perturbations are not ``forgotten'' in the evolution (apart from
$\hat{\bar v}_{ci}$). We shall take (rather arbitrarily)
\begin{equation}
\label{ci_res_fin}
\hat{\bar v}^i_{n{\rm in}}=0.
\end{equation}

In conclusion, we shall start the numerical integration of the set
(\ref{pert_nu}-\ref{pert_tens}) with the initial conditions
(\ref{ci_res_deb}-\ref{ci_res_fin}), at $x=x_{\rm in}$ (in practice $x_{\rm
in}=10^{-9}$), for various sets of the constants $f_2$, $f_4$, $f_6$
and $f_7$ (which will depend on our choices for the functions
${\mathcal F}(u)$, ${\mathcal G}(u)$, ${\mathcal P}(u)$ and ${\mathcal
Q}(u)$ which describe the defects [see
\EQNS{\ref{jpu_f1}-\ref{jpu_f7}}]), and all relevant values of $q$
(in practice $5\times10^{-2}<q<5\times10^2$, which corresponds to
wavelengths from 10 times to one thousandth of the horizon size
today).

An important point to note is that all scalar perturbations,
collectively called $\hat S(\eta, k^i)$, are proportional to the same
random variable $e(k^i)$ such that $\langle e^*(k'^i)e(k^i)\rangle$
$=\delta(k'^i-k^i)$, and can be written as
\begin{equation}
\label{res_coh_scal}
\hat S(k^i)=\tilde S(k)e(k^i)
\end{equation}
where $\tilde S$ is a function of $k$ only which is entirely
determined once the two functions $f_1(u)$ and ${\mathcal G}(u)$
describing the scalar components of the defects are known. As for the
vector perturbations, collectively called $\hat{\bar V}_i(k^i)$, they
can be written as
\begin{equation}
\label{res_coh_vec}
\hat{\bar V}_i(k^i)=\tilde{\bar V}(k)\bar e_i(k^i)
\end{equation}
with [see (\ref{corr_coh_v})] $\langle \bar e_i^*(k'^i)\bar
e_k(k^i)\rangle=\delta(k'^i-k^i)P_{ik}$. Finally the tensor
perturbations, collectively called $\hat{\bar T}_{ij}(k^i)$, can be
written as
\begin{equation}
\label{res_coh_tens}
\hat{\bar T}_{ij}(k^i)=\tilde{\bar T}(k)\bar e_{ij}(k^i)
\end{equation}
with [see (\ref{corr_coh_t})] $\langle \bar e_{ij}^*(k'^i)\bar
e_{kl}(k^i)\rangle
=\delta(k'^i-k^i)(P_{ik}P_{jl}+P_{il}P_{jk}-P_{ij}P_{kl})$.  The
functions $\tilde{\bar V}(k)$ and $\tilde{\bar T}(k)$ are entirely
determined once the functions ${\mathcal P}(u)$ and ${\mathcal Q}(u)$
describing the vector and tensor components of the defects is known.

\section{The correlation function of the microwave background anisotropies}
\label{sec_cmb}
 
The energy of a microwave background photon in the direction
$\gamma^i$, as measured by an observer $O$ at $x^\mu_0$ is~: $E^0_{\vec
\gamma}=-(g_{\mu\nu}K^\mu u^\nu)|_0$ where $g_{\mu\nu}|_0$ is the
metric (\ref{frw_pert}) evaluated at $x^\mu_0$, $K^\mu|_0$ (with
$g_{\mu\nu}K^\mu K^\nu=0$) is the wave vector of the photon at
$x^\mu_0$ and $u^\nu|_0$ (with $g_{\mu\nu}u^\mu u^\nu=-1$) is the
velocity vector of the observer $O$.  $E^0_{\vec \gamma}$ depends on
$K^0|_0$, on the direction of observation $\gamma^i\equiv-(K^i/|K|)_0$
and the position and velocity of the observer $O$.  Assuming that the
photon has followed a nul geodesic (no-collision hypothesis) since it
has decoupled from the baryons on a surface of constant electronic
density or, almost equivalently, constant photon density, it is a
standard calculation (see \EG{} \cite{panek} for a detailed
derivation) to relate $E^0_{\vec \gamma}$ to $E^d=-(g_{\mu\nu}K^\mu
u_b^\nu)|_{\rm d}$, its energy at decoupling, which is independent of
$\gamma^i$ when ``measured'' by an observer comoving with the baryons.
The result is
\begin{eqnarray}
\label{dtt}
\left({\Delta T\over T}\right)_{\vec\gamma} \equiv  
{E^0_{\vec \gamma}-\bar E^0\over\bar E^0}
 & = & -\left[\Phi+{1\over4}\delta_\gamma
              -\gamma^i\left(\partial_iv+\bar v_i\right)\right]_0 \\
\nonumber &   & +\left[\Phi+{1\over4}\delta_\gamma
              -\gamma^i\left(\partial_iv_b+\bar v_{i, b}\right)
                       \right]_{\rm d} \\
\nonumber &   & +\int_{\eta_d}^{\eta_0} 
        \left(\Phi'+\Psi'-\gamma^i\bar\Phi_i'-\gamma^i\gamma^j\bar
E_{ij}'\right)\ddd\eta
\end{eqnarray}
where $\bar E^0\equiv E^dx_{\rm d}$ ($x_{\rm d}\equiv
a(\eta_d)/a(\eta_0)$), where the index $0$ means that the quantity is
evaluated today at ($\eta_0, x^i_0=0$), where the index $d$ means that
the quantity is evaluated at ($\eta_d, x^i_d=\gamma^i(\eta_0-\eta_d)
$) and where the integrands in the integrals are evaluated on the
photon trajectory that is at $x^\mu=\left(\eta,
\gamma^i(\eta_0-\eta)\right)$.  Assuming that the microwave background
is and always was a perfect black body radiation,
\EQ{\ref{dtt}} defines its temperature anisotropies.

\EEQ{\ref{dtt}}, where the reception event $x^\mu_0$ is defined
(rather arbitrarily) as belonging to a surface of constant photon
density, defines $\Delta T/T|_{\vec\gamma}$ as a gauge invariant
quantity. The first term is the sum of a monopole
($-[\Phi+{1\over4}\delta_\gamma]_0$) which does not depend on
$\gamma^i$ and a dipole ($\left[\gamma^i\left(\partial_iv+\bar
v_i\right)\right]_0$) which is a Doppler effect due to the peculiar
velocity of the observer. In the following we shall ignore these
contributions and define the temperature background anisotropies as
\begin{equation}
\Theta_{\vec\gamma} = 
  \Theta_{\vec\gamma}^S+\Theta_{\vec\gamma}^V+\Theta_{\vec\gamma}^T
\end{equation}
with 
\begin{eqnarray}
\label{dtt_s}
\Theta_{\vec\gamma}^S & = & 
\left[\Phi+{1\over4}\delta_\gamma-\gamma^i\partial_iv_b\right]_{\rm d}
+\int_{\eta_d}^{\eta_0}\!\left(\Phi'+\Psi'\right)\,\ddd\eta, \\
\Theta_{\vec\gamma}^V & = & 
-\gamma^i\left(\bar v_{i, b}|_{\rm d}
               +\int_{\eta_d}^{\eta_0}\!\bar\Phi_i'\,\ddd\eta\right), \\
\label{dtt_t}
\Theta_{\vec\gamma}^T & = & 
-\gamma^i\gamma^j\int_{\eta_d}^{\eta_0}\!\bar E_{ij}'\,\ddd\eta.
\end{eqnarray}
The next step consists in expressing the perturbations in terms of
their Fourier transforms [see \EQ{\ref{TF}}] and to expand
$e^{{\rm i}k_ix^i}$ as
\begin{equation}
e^{{\rm i}k_ix^i} = 
  \sum_\ell {\rm i}^\ell(2\ell+1)j_\ell(kx)P_\ell(\vec k.\vec x/kx)
\end{equation}
where $j_\ell$ is a Bessel function, $P_\ell$ a Legendre polynomial,
and where $kx\equiv\sqrt{k_ik^i}\sqrt{x_ix^i}$ and $\vec k.\vec
x\equiv k_ix^i$.  \EEQNS{\ref{dtt_s}-\ref{dtt_t}} then read
(reintroducing the dimensionless wave vector $q^i\equiv k^i/{\mathcal
H}_0$)
\begin{equation}
\label{gkl_svt}
\Theta_{\vec\gamma}^{S,V,T}={1\over(2\pi)^{3/2}}\sum_\ell{\rm
i}^\ell(2\ell+1)\int \! q^2\ddd q\, \ddd\Omega_{\vec q}\, 
P_\ell(\vec n_q.\vec\gamma)\, G^{S,V,T}_\ell(q^i)
\end{equation}
where $S,V,T$ stand for either $S$, $V$ or $T$, where $\Omega_{\vec
q}$ defines the direction of the vector $q^i$ in Fourier space, where
$\vec n_q \equiv \vec q / q$ and where [integrating on $x\equiv a/a_0$
rather than on $\eta$ and recalling the notations
(\ref{res_coh_scal}-\ref{res_coh_tens})]
\begin{eqnarray}
G^S_\ell(q^i) & \equiv & \tilde G^S_\ell(q)e(q^i), \\
G^V_\ell(q^i) & \equiv & \tilde G_\ell^V(q)\gamma^i\bar e_i(q^i), \\
G^T_\ell(q^i) & \equiv & \tilde G_\ell^T(q)\gamma^i\gamma^j\bar e_{ij}(q^i),
\end{eqnarray}
with
\begin{eqnarray}
\label{glk_s}
\tilde G^S_\ell(q) & = & 
 j_\ell(y_{\rm d})\left[\tilde\Phi+{1\over4}\tilde\delta_\gamma\right]_{\rm d}
 -\left[\frac{\ddd}{\ddd y}j_\ell(y)\right]_{\rm d} \tilde V_b|_{\rm d}\\
\nonumber&&
 +\int_{x_{\rm d}}^1\!\ddd x\, 
  j_\ell(y)\frac{\ddd}{\ddd x}(\tilde\Phi+\tilde\Psi), \\
\tilde G_\ell^V(q) & = & 
 -j_\ell(y_{\rm d})\tilde{\bar v}_{b}|_{\rm d}
 -\int_{x_{\rm d}}^1\!\ddd x\, j_\ell(y)
  \frac{\ddd}{\ddd x}\tilde{\bar \Phi}, \\
\label{glk_t}
\tilde G_\ell^T(q) & = & 
 -\int_{x_{\rm d}}^1\!\ddd x\,j_\ell(y)\frac{\ddd}{\ddd x}\tilde{\bar E},
\end{eqnarray}
where $y\equiv
kr=k(\eta_0-\eta)=2q(1-\sqrt{\Omega_r^0+x\Omega_m^0})/\Omega_m^0$ [see
\EQ{\ref{eta}}] and where the index ``${\rm d}$'' means that the
quantity is evaluated at $x=x_{\rm d}$.

$\Theta_{\vec\gamma}^S$, $\Theta_{\vec\gamma}^V$, and
$\Theta_{\vec\gamma}^T$ are random variables which fall into
statistically independent sets so that the temperature anisotropy
correlation function factorizes as
\begin{equation}
\label{c_theta}
C(\vec\gamma_1.\vec\gamma_2)\equiv
\langle\Theta_{\vec\gamma1}\Theta_{\vec\gamma2}
\rangle=\langle \Theta^S_{\vec\gamma1}\Theta^S_{\vec\gamma2}\rangle +\langle
\Theta^V_{\vec\gamma1}\Theta^V_{\vec\gamma2}\rangle+\langle
\Theta^T_{\vec\gamma1}\Theta^T_{\vec\gamma2}\rangle
\end{equation}
where $\vec\gamma_1.\vec\gamma_2 $ is the cosine of the angle between
the two directions of observation $\gamma_1^i$ and $\gamma_2^i$.

Injecting \EQNS{\ref{gkl_svt}-\ref{glk_t}} into \EQ{\ref{c_theta}}
one therefore gets
\begin{eqnarray}
\langle \Theta_{\vec\gamma_1}^{S,V,T}\Theta_{\vec\gamma_2}^{S,V,T}\rangle &=&  
{1\over(2\pi)^3}\sum_{\ell,\ell'}
(-{\rm i})^{\ell'}{\rm i}^\ell(2\ell'+1)(2\ell+1) \times\\
\nonumber&\int&
q^2\ddd q\,\ddd\Omega_{\vec q}\, 
\tilde G^{*S,V,T}_{\ell'}\tilde G^{S,V,T}_\ell 
P_{\ell'}(\vec n_q.\vec\gamma_1)P_\ell(\vec n_q.\vec\gamma_2)
{\mathcal A}^{S,V,T},
\end{eqnarray}
where
\begin{eqnarray}
{\mathcal A}^S & = & 1, \\
{\mathcal A}^ V&=&\gamma^i_2\gamma^j_1 P_{ij}  =  
\vec\gamma_1.\vec\gamma_2-(\vec n_q.\vec\gamma_1)(\vec n_q.\vec\gamma_2),\\
{\mathcal A}^T &=& \gamma_2^i\gamma_2^j\gamma_1^k\gamma_1^l
(P_{ik}P_{jl}+P_{il}P_{jk}-P_{ij}P_{kl}) \\
\nonumber&=& 
  2[\vec\gamma_1.\vec\gamma_2
    -(\vec n_q.\vec\gamma_1)(\vec n_q.\vec\gamma_2)]^2
  -[1-(\vec n_q.\vec\gamma_1)^2][1-(\vec n_q.\vec\gamma_2)^2].
\end{eqnarray}
Using now the identity \cite{gradstein}
\begin{equation}
\int\! \ddd\Omega_{\vec q}\,P_\ell(\vec n_q.\vec\gamma_1)\times P_{\ell'}(\vec
n_q.\vec\gamma_2) ={4\pi\over
2\ell+1}P_\ell(\vec\gamma_1.\vec\gamma_2)\delta_{\ell\ell'}
\end{equation}
one first obtains (see \EG{} \cite{abbot})
\begin{eqnarray}
\langle \Theta_{\vec\gamma_1}^S\Theta_{\vec\gamma_2}^S\rangle & = &
  {1\over4\pi}\sum_\ell(2\ell+1)C_\ell^S P_\ell(\vec\gamma_1.\vec\gamma_2), \\
\label{c_ell_s}
C_\ell^S & = & {2\over\pi}\int q^2\ddd q|\tilde G_\ell^S|^2.
\end{eqnarray}
where $\tilde G_\ell^S$ is the function of $q$ given by
\EQ{\ref{glk_s}}.  The other identity \footnote{We thank Luc
Blanchet for showing us how to demonstrate it by means of the STF
formalism (see \EG{} \cite{luc})}
\begin{eqnarray}
\int (\vec n_q.\vec\gamma_1) P_\ell(\vec n_q.\vec\gamma_1)
     (\vec n_q.\vec\gamma_2) P_{\ell'}(\vec n_q.\vec\gamma_2) 
     \ddd\Omega_{\vec q} \\ 
\nonumber
=4\pi{\ell+1\over(2\ell+3)(2\ell+1)}
 \left[{\ell+1\over2\ell+1}\delta_{\ell\ell'}
       +{\ell+2\over2\ell+5}\delta_{\ell',\ell+2}\right]
 P_{\ell+1}(\vec\gamma_1.\vec\gamma_2) \\
\nonumber
 +  4\pi{\ell\over(2\ell+1)(2\ell-1)}
\left[{\ell\over 2\ell+1}\delta_{\ell\ell'}
      +{\ell-1\over2\ell-3}\delta_{\ell',\ell-2}\right]
 P_{\ell-1}(\vec\gamma_1.\vec\gamma_2) 
\end{eqnarray}
together with standard recursion relations for the Legendre
polynomials and Bessel functions \cite{gradstein} then yield (see
\EG{} \cite{pert_gen})
\begin{eqnarray}
\langle\Theta_{\vec\gamma_1}^V\Theta_{\vec\gamma_2}^V\rangle
& = & {1\over4\pi}\sum_\ell(2\ell+1)C_\ell^V
P_\ell(\vec\gamma_1.\vec\gamma_2), \\
\label{c_ell_v}
C_\ell^V & = & {2\over\pi}\ell(\ell+1)\int q^2\ddd q
  \left[{j_\ell(y_{\rm d})\over y_{\rm d}}\tilde{\bar v}_{b}|_{\rm d}
        +\int_{x_{\rm d}}^1\!\ddd x\, {j_\ell(y)\over y}
         \frac{\ddd}{\ddd x}\tilde{\bar \Phi}
  \right]^2.
\end{eqnarray}

The computation of $\langle
\Theta_{\vec\gamma_1}^T\Theta_{\vec\gamma_2}^T\rangle$ can proceed
along similar lines but is more involved. The result is (see
\EG{} \cite{abbot})
\begin{equation}
\label{c_ell_t}
\langle \Theta_{\vec\gamma_1}^T\Theta_{\vec\gamma_2}^T\rangle =
{1\over4\pi}\sum_\ell(2\ell+1)C_\ell^T
P_\ell(\vec\gamma_1.\vec\gamma_2)
\end{equation}
with
\begin{equation}
C_\ell^T = \frac{2}{\pi}\frac{(\ell+2)!}{(\ell-2)!}
\int q^2 \ddd q
\left[\int_{x_{\rm d}}^1\!\ddd x\, \frac{j_\ell(y)}{y^2}
\frac{\ddd}{\ddd x}\tilde{\bar E}\right]^2.
\end{equation}

The coefficients $C_\ell^{S,V,T}$, which will be computed numerically
in the next section for various choices of the source functions
describing the defects, are the quantities to be compared with
observations, through the inverse formula
\begin{eqnarray}
C_\ell^S+C_\ell^V+C_\ell^T & = & 2\pi\int_{-1}^{+1}\!\ddd z\,C(z) P_\ell(z), \\
z    & \equiv & \vec\gamma_1.\vec\gamma_2, \\ 
C(z) & \equiv & \langle \Theta_{\vec\gamma_1}\Theta_{\vec\gamma_2}\rangle.
\end{eqnarray}
(Observationally, the ensemble average $\langle...\rangle$ is replaced
by an average on different regions of the celestial sphere. For large
angles there are few such independent regions so that the ``cosmic
variance'' problem arises~; see \EG{} \cite{bunn}.)

\section{Numerical results}
\label{sec_res}

Two of us (JPU and AR) have developped a numerical code to solve the
equations for the perturbations (\ref{pert_nu}-\ref{pert_tens}) and compute
the coefficients $C_\ell$ in function of the multipoles $\ell$
[\EQNS{\ref{c_ell_s},\ref{c_ell_v},\ref{c_ell_t}}] when the defects are
described by \EQNS{\ref{corr_coh_rho}-\ref{corr_coh_pit}} and
(\ref{jpu_f1}-\ref{jpu_f7}), and the initial conditions are given by
\EQNS{\ref{ci_res_deb}-\ref{ci_res_fin}} (the computations start at
$x_{\rm{in}}=10^{-9}$, $q\in[2\times10^{-2}-2\times 10^{+3}]$ and
$\ell\in [2-1500]$).  This code incorporates Silk damping (that is the
decay of the $C_\ell$ at large $\ell$ due to the fact that a fluid
description no longer holds on small scales) by a simple
multiplication of $G^S$ in (\ref{glk_s}) by the decaying exponential
(see \cite{kaiser}-\cite{fusuun})
\begin{equation}
\label{damping}
{\rm e}^{-2k^2\eta^2_{\rm rec}x_s^2},
\end{equation} 
where $x_s=0.6\Omega_m^{1/4}(\Omega_b h_0)^{-1/2}a_{\rm rec}^{3/4}$
\cite{seljak94}.

As for the fact that decoupling is not instantaneous, we take it into
account by convolving \EQNS{\ref{glk_s}-\ref{glk_t}} with the
visibility function (see \cite{pert_gen} for a detailed derivation; a
simple multiplication by ${\rm e}^{-k^2\eta_{\rm rec}^2\sigma^2}$ with
$\sigma=0.03$ for standard recombination \cite{seljak94}, is too rough an
approximation).

Instead of \EQ{\ref{damping}} we could have used the better fit
proposed by Hu and Sugiyama \cite{husu} but it is an analytical fit
deduced from a Boltzmann code whereas \EQ{\ref{damping}} is obtained
from physical arguments \cite{kaiser}.

In this section we first test our code and discuss the approximations
made. We then give a few examples of solutions and compare the results
with previous calculations. (Note that for a better visibility, the
results are not normalized to the COBE data at $\ell=10$ as is usual
\cite{tegmark}.)

\subsection{No sources and adiabatic conditions~: comparison with a
Boltzmann code}

The code which we have implemented can of course also be used to study
scenarios where no defects are present and where adiabatic initial
conditions (as imposed by \EG{} the standard inflationary scenario)
are imposed. One has just to set ${\mathcal F}={\mathcal G}={\mathcal
P}={\mathcal Q}=0$ and replace the initial conditions
(\ref{ci_res_deb}-\ref{ci_res_fin}) by
\begin{eqnarray}
\hat \delta^\flat_\nu=\hat
\delta^\flat_\gamma={4\over3}\hat \delta^\flat_b= {4\over3}\hat
\delta^\flat_c & = & -6\Psi_0 \\
\hat V_\nu=\hat V_\gamma=\hat V_b=\hat
V_c & = & -{1\over2}{qx\over\sqrt{\Omega_r^0}}\Psi_0\\
\hat\Psi=\hat\Phi& = & \Psi_0
\end{eqnarray}
where $\Psi_0$, in the standard chaotic inflationary model, is fixed
by the quantization of the inflaton so that (see \EG{} \cite{linde})
$\Psi_0\propto q^{-3/2}$. (We restrict ourselves to the scalar
perturbations as the vector and tensor contributions are negligible in
the standard inflationary scenario --- see \EG{} \cite{nath}.)

\begin{figure}
\centering
\psfig{file=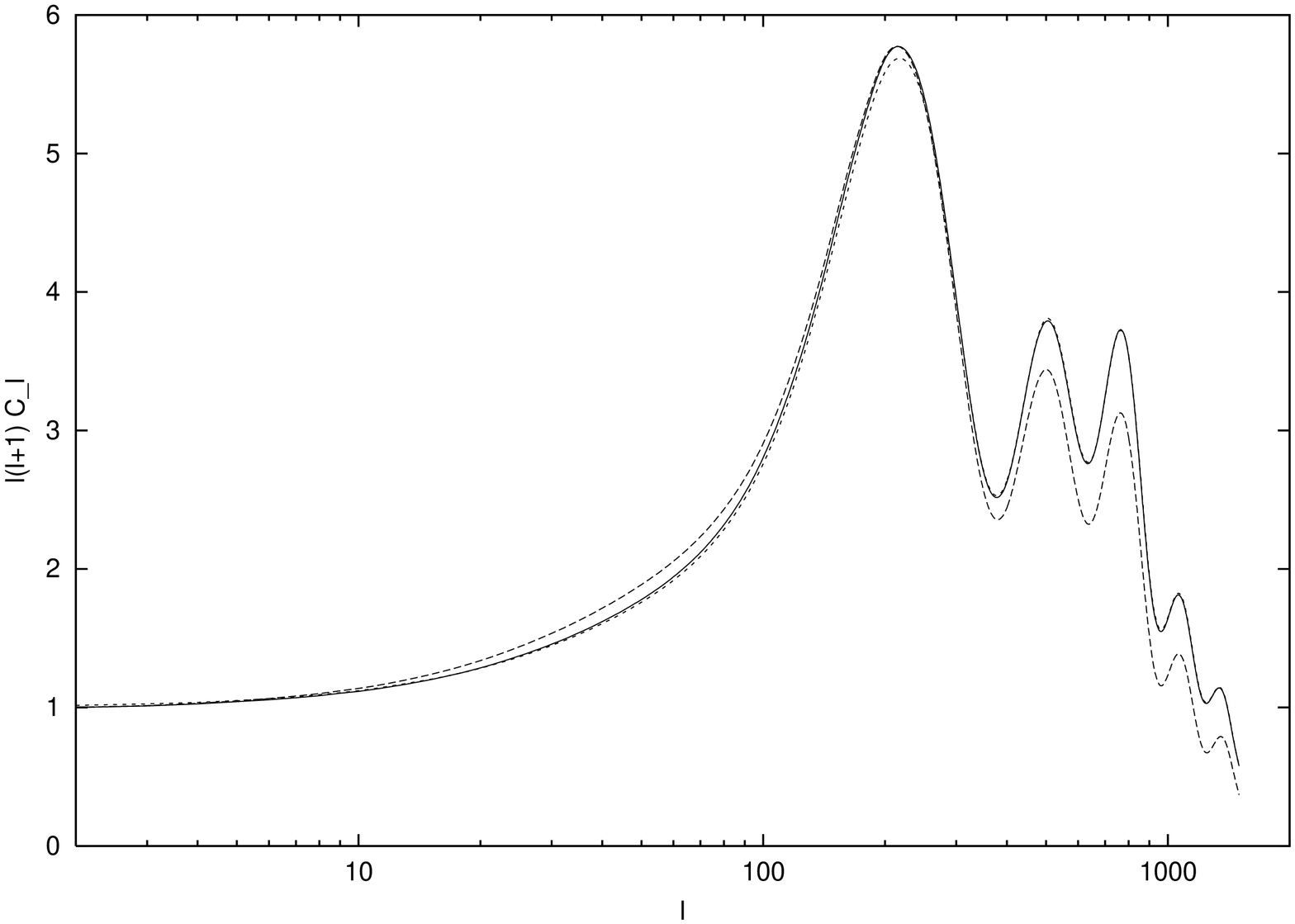,width=3.1in,angle=-90}
\psfig{file=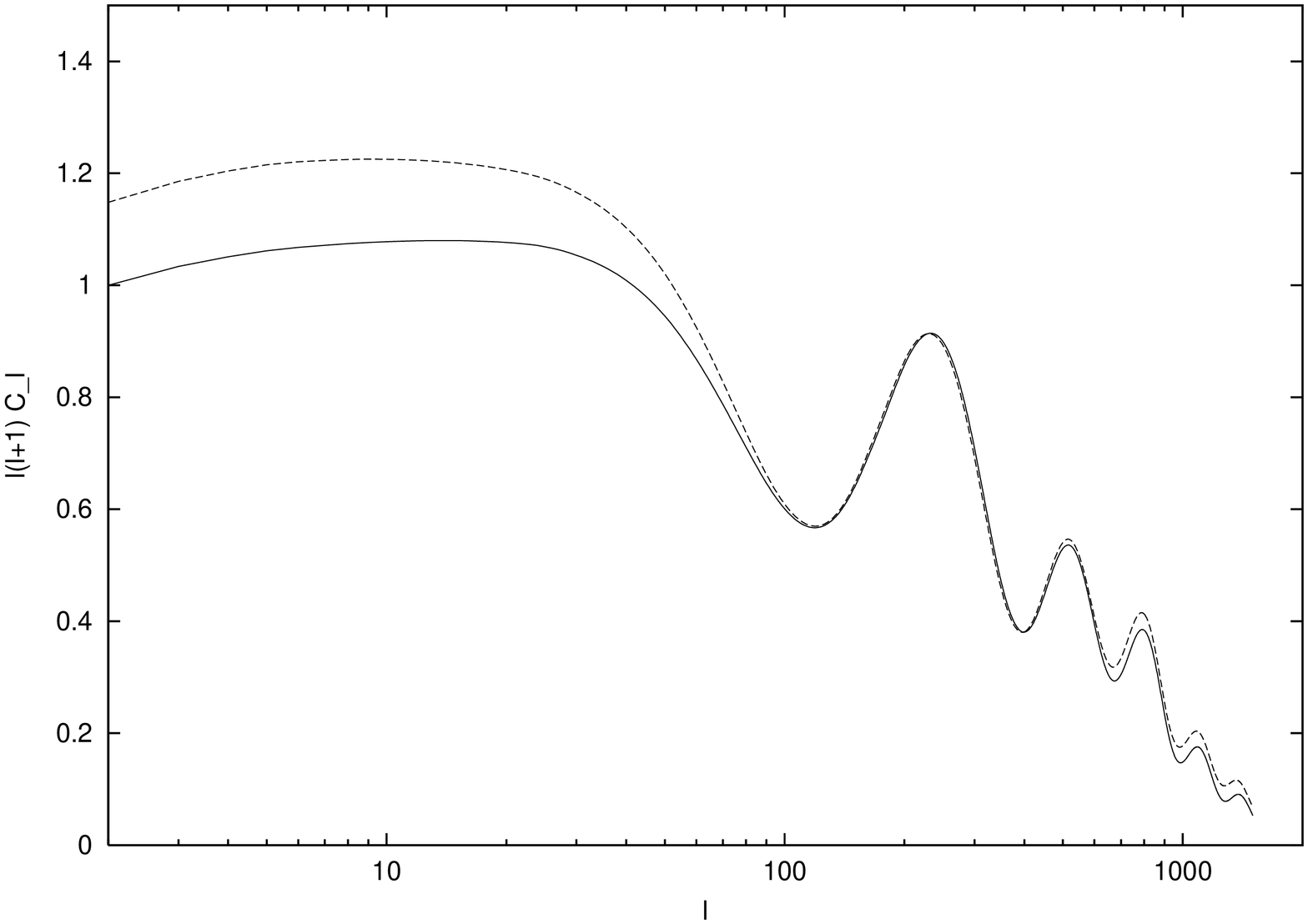,width=3.1in,angle=-90}
\caption{Comparison between our fluid code (solid line), our Boltzmann
code (dashed line) and CMBFAST (short-dashed line) in the case of inflation
(top) ; Comparison between our fluid code (solid line) and our
Boltzmann code (dashed line) in the case of topological defects
(bottom) with ${\mathcal F}=\exp(-u^2)$ and ${\mathcal G}=0$.}
\label{fig1}
\end{figure}

The $C^S_\ell$ coefficients as computed with our fluid code, with the
Boltzmann code we have developped \cite{rdxx}, and with CMBFAST
\cite{cmbfast_code} are compared on \FIG{\ref{fig1}} (top). One first
notes the excellent agreement between our Boltzmann code with CMBFAST.
One also notes that if the position of the Doppler peaks, as given by
our fluid code and the Boltzmann ones, coincide, their heights differ
by about 15 \%, which we consider enough for our present purposes
considering that the physics in our fluid code (as well as in \EG{}
Seljak's in
\cite{seljak94}) and a Boltzmann code differs on four points~:
\begin{itemize}
  \item We use a perfect fluid approximation to describe the various
  matter and radiation components of the universe,

  \item We make a tight-coupling approximation, that is we assume that
  the baryons and photons have exactly the same velocities until
  decoupling,

  \item We suppose instantaneous decoupling,

  \item We correct for those assumptions by an adhoc damping of the
  scalar perturbations on small scales [see \EQ{\ref{damping}}] and we
  take into account the finite width of the last scattering surface.
\end{itemize}
whereas a description based on Boltzmann's equation is used in our
Boltzmann code and in CMBFAST. We find that the difference in the
height of the peaks is due to these approximations (see \cite{rdxx}
for a detailed discussion).

In conclusion a perfect fluid code is useful to find the {\sl
positions} of the peaks predicted by various types of topological
defects but can be inaccurate when it comes to their {\sl height}. In
the following we shall use a perfect fluid code and therefore comment
only on the position of the peaks. In \cite{rdxx} we shall use our
Boltzmann code and shall therefore be able to compare not only the
positions but also the heights of the peaks predicted by various
defects as well as with those predicted by inflationary scenarios, the
latter being also computed using a Boltzmann code.

\subsection{The scalar contribution of coherent defects to the CMB
anisotropies}

The scalar component $C_\ell^S$ of the $C_\ell$ coefficients is given
by (\ref{c_ell_s}) and is entirely determined once the two functions
${\mathcal F}(u)$ and ${\mathcal G}(u)$ (with $u\equiv q/h=k/{\mathcal
H}$) describing the scalar component of the stress-energy tensor of
the defects are known.

\begin{figure}
\centering
\psfig{file=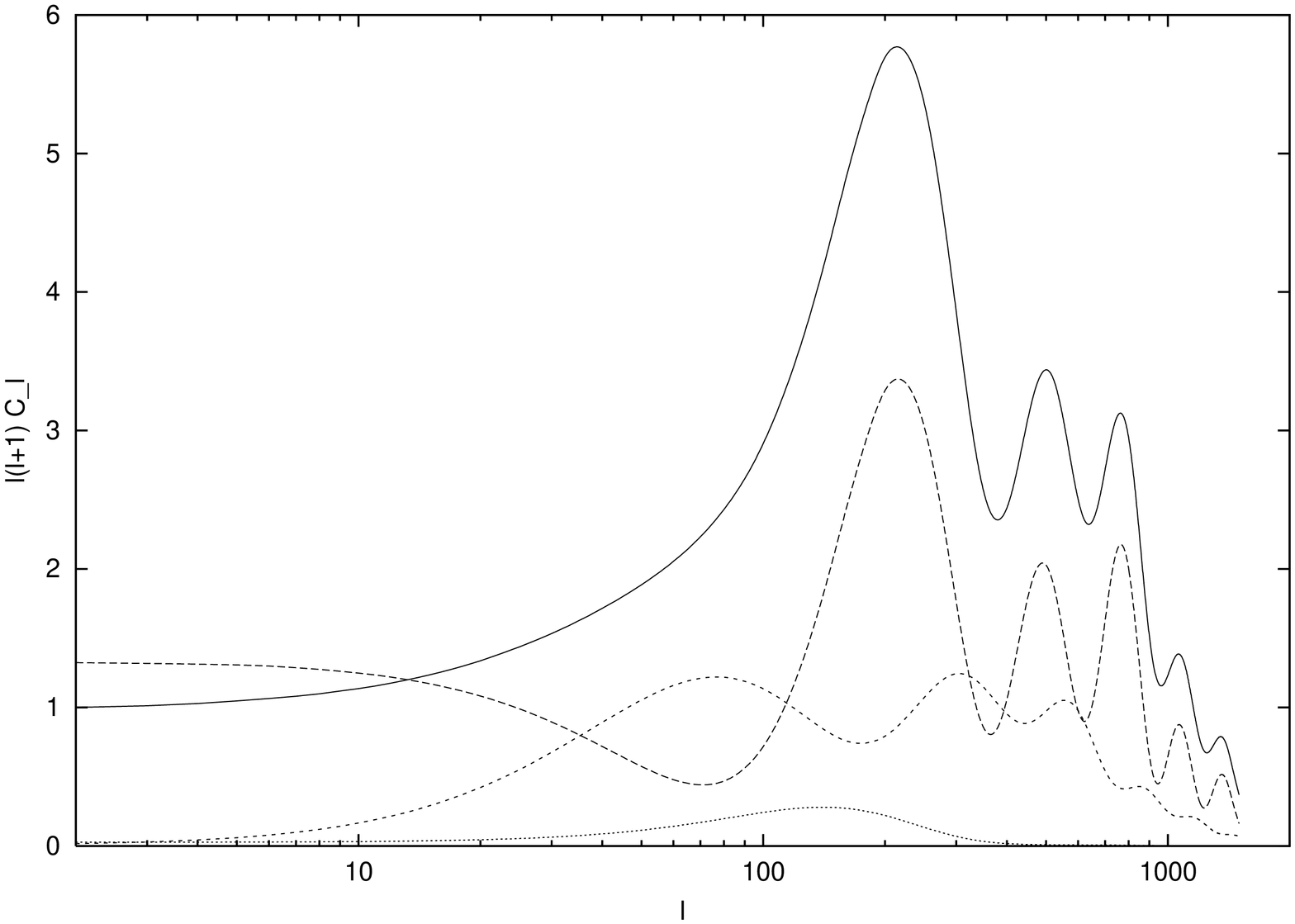,width=3.1in,angle=-90}
\psfig{file=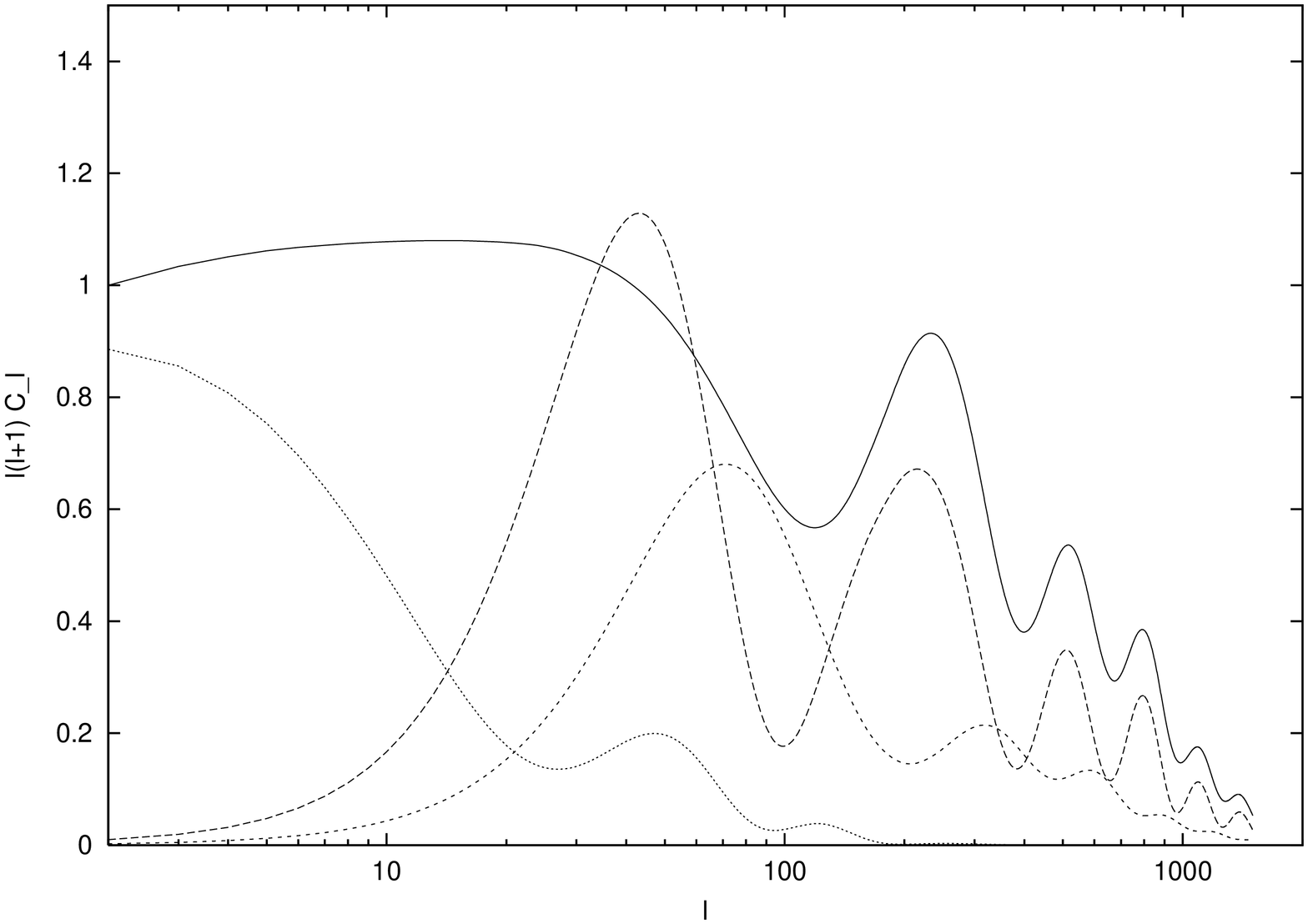,width=3.1in,angle=-90}
\caption{Decomposition of the different contributions of the
$C^S_\ell$ in the case of inflation (top) and topological defects
(bottom) with ${\mathcal F}=\exp(-u^2)$ and ${\mathcal G}=0$. In
each case, we have represented the Sachs-Wolfe, Doppler, Integrated
Sashs-Wolfe contribution, as well as the total contribution (dashed,
short-dashed, dotted and solid lines respectively).}
\label{fig2}
\end{figure}

First \FIG{\ref{fig2}} shows, on the particular case ${\mathcal
G}(u)=0$ and ${\mathcal F}(u)\propto \exp(-u^2)$ (\FIG{\ref{fig2}},
bottom), that, contrarily to what happens in inflationary scenarios
(\FIG{\ref{fig2}}, top), the integrated Sachs-Wolfe (ISW) contribution
to $C^S_\ell$ (that is the integral term in \EQ{\ref{glk_s}})
dominates at low $\ell$. This is in agreement with intuition~: in
inflationary scenarios the large scale anisotropies are built in the
initial conditions, whereas in defect scenarios they are due to
line-of-sight gravitational effects. We are aware that the inverse
Fourier transform of a gaussian does not meet condition
(\ref{corr_causal}) but we note that the Fourier transform of any
compact supported $C^\infty$ function decreases faster than any power
law.

\begin{figure}
\centering
\psfig{file=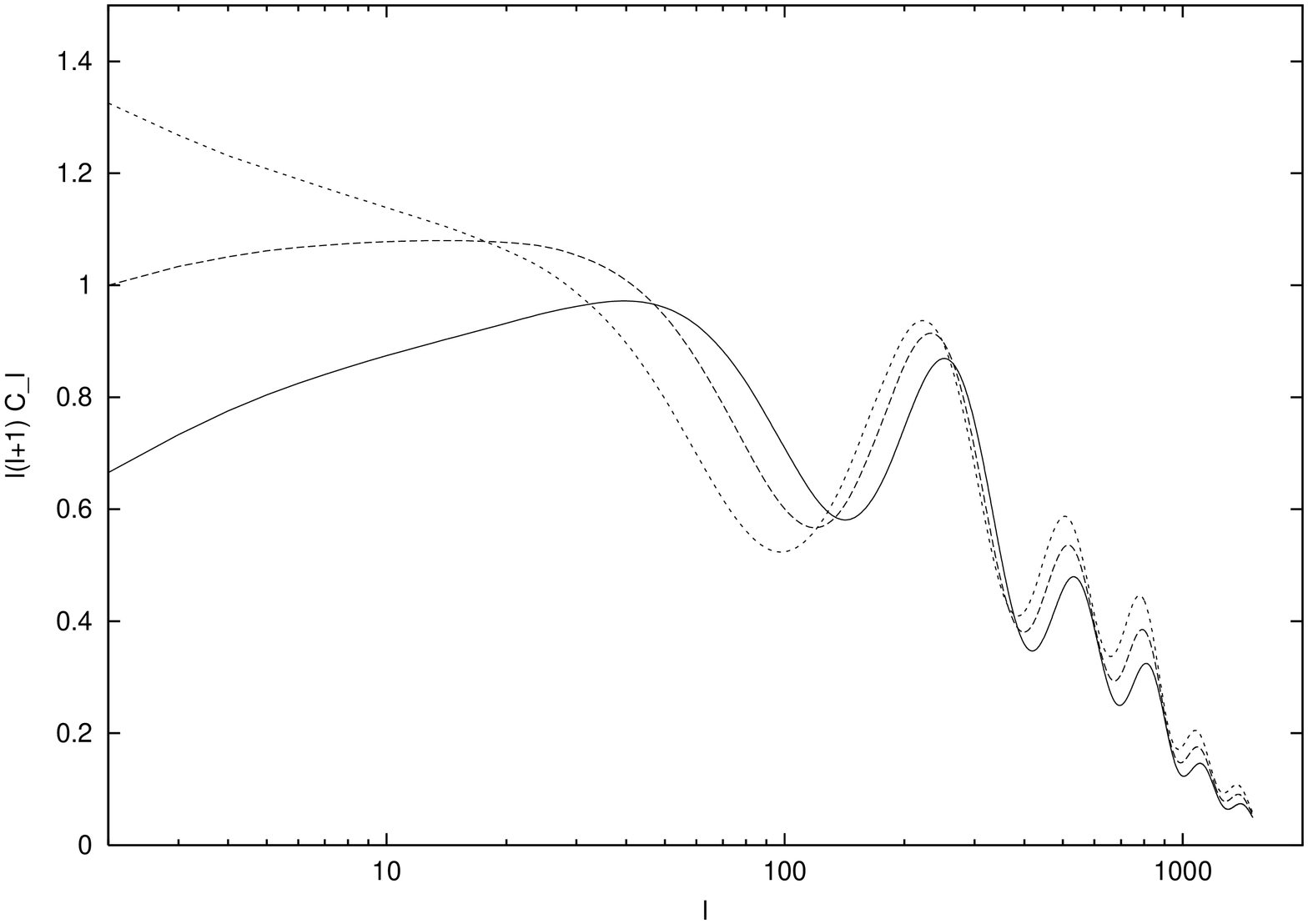,width=3.1in,angle=-90}
\caption{The dependence of the coefficients $C^S_\ell$ in defect scenario
where we have varied the scale of decay, $L$, of the defects by
choosing ${\mathcal F}=\exp(-u^2/L^2)$ and ${\mathcal G}=0$, with $L^2
= 0.5, 1, 2$ (dotted, dashed and solid line respectively).}
\label{fig3}
\end{figure}

\FFIG{\ref{fig3}} illustrates the case when ${\mathcal G}=0$ and
${\mathcal F}= \exp(-u^2/L^2)$ for various values of $L$.  The most
striking feature of these curves is that they do not exhibit a plateau
at low $\ell$ when $L\neq 1$. As $\ell$ decreases the first ``peak''
is shifted to lower multipoles and tends to the position predicted by
the standard inflationary scenario (see \FIG{\ref{fig1}}, top). This
can be explained by the fact that the defects decay on larger scales.

\begin{figure}
\centering
\psfig{file=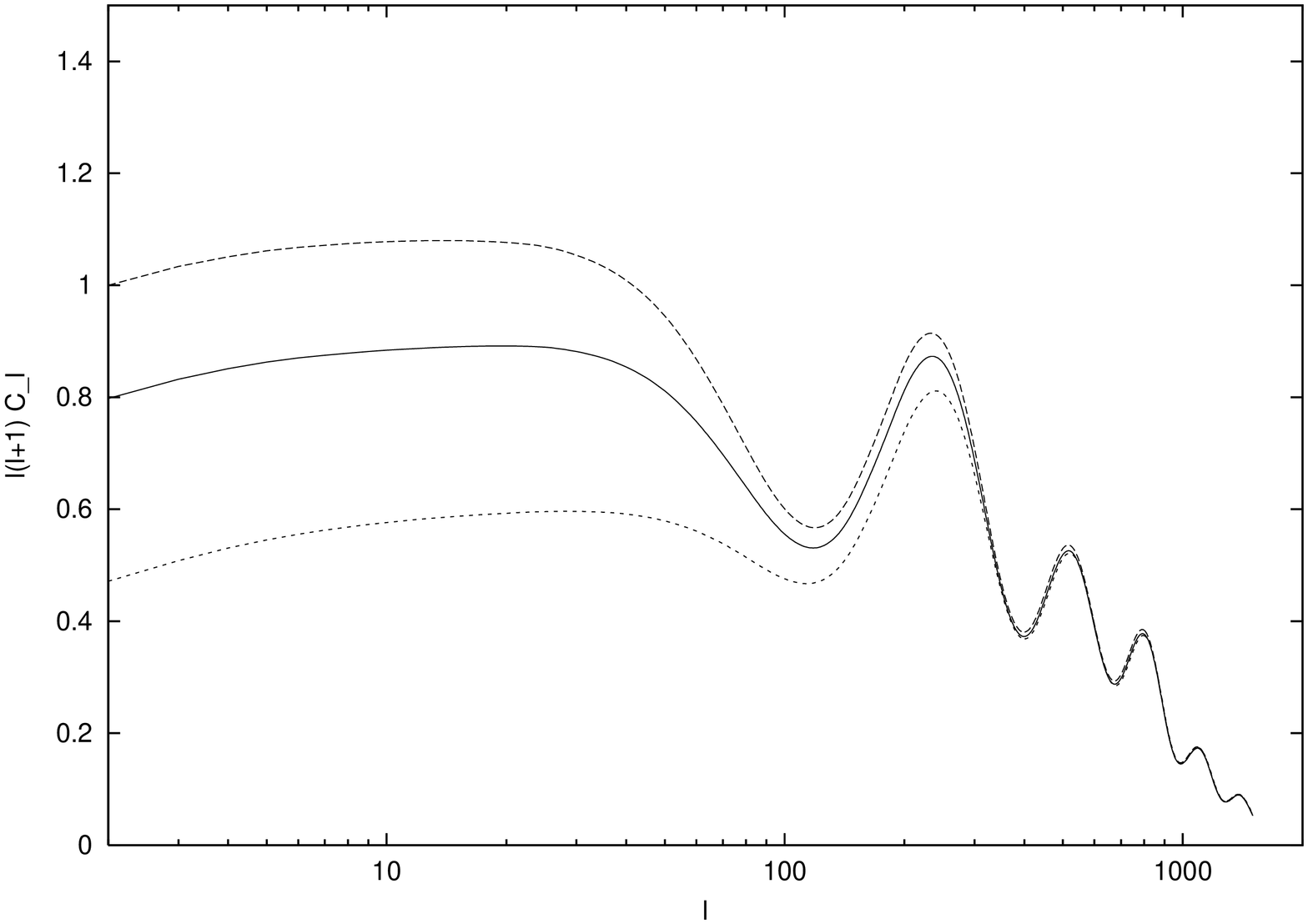,width=3.1in,angle=-90}
\caption{The effect of the function ${\mathcal G}=\exp(-u^2)$. 
For that purpose we have varied the ratio $r={\mathcal G}/{\mathcal
F}$ with $r=0, 0.1, 0.3$ (dotted, dashed and solid line
respectively).}
\label{fig4}
\end{figure}

\FFIG{\ref{fig4}} illustrates the case when both ${\mathcal F}$ and
${\mathcal G}$ are proportional to $\exp(-u^2)$, with ${\mathcal
G}/{\mathcal F}=r$. First, as expected, when $r$ is small one recovers
\FIG{\ref{fig2}}, bottom or \FIG{\ref{fig3}} with $L=1$.  Second, when
${\mathcal G}\neq0$, the dominant component to the stress-energy
tensor of the defects is their anisotropic stress $\Pi^s$ [see
\EQNS{\ref{jpu_f1}-\ref{jpu_f4} and \ref{corr_coh_rho}}]. The first
Doppler peak is then washed out. This result compares to \cite{pst1}
where incoherent sources are studied, for which the anisotropic stress
dominates too. An interesting particular case is when $a\simeq0.3$
which minimizes the anisotropic stress contribution $f_4$ [see
\EQ{\ref{jpu_f4}}].  The first Doppler peak is then higher than the
plateau.

Finally, as a caveat not to comment too seriously on the height of the
Doppler peaks when a perfect fluid code is used, \FIG{\ref{fig1}},
bottom compares $C_\ell^S$ as given by the perfect fluid code used here
and the Boltzmann code used in \cite{rdxx} for ${\mathcal G}(u)=0$ and
${\mathcal F}(u)\propto \exp(-u^2)$. The much bigger discrepancy in the
height of the peaks in that case as compared to the case of inflation
is due to the fact that \EQ{\ref{damping}} for Silk damping is valid
only if the metric perturbations in the evolution equations for
$\delta_\gamma$ can be aproximated by constant (see \EG{} \EQ{98} of
Hu and White \cite{pert_gen}). It can be shown that this is
approximatively true in the case of inflationnary scenarios in a flat
universe without cosmological constant, but here it is not true, as it
can be checked by looking at the ISW contribution. Of course, it is
also possible to do some a posteriori corrections by comparing fluid
and Boltzman codes and by fitting ad hoc damping terms in order to
find a good agreement between the two, see Hu and Sugiyama
\cite{pert_gen}, but we did not find them necessary since we simply
wanted to present a semi-analytic formalism. Let us note finally that
in the case of the pressure model \cite{huspwh}\cite{cheungma}, the
discrepancy between this (simple) fluid code and an accurate Boltzmann
code is less serious, probably because the ISW contribution is
smaller.

\subsection{The vector and tensor contributions}

The vector and tensor components of the $C_\ell$ coefficients [given
by \EQNS{\ref{c_ell_v},\ref{c_ell_t}}] are entirely determined once the
functions ${\mathcal P}(u)$ and ${\mathcal Q}(u)$ describing the
vector and tensor components of the stress-energy momentum of the
defects are known. Since we supposed $\bar v_b^i=0$ they both reduce
to their integrated Sachs-Wolfe contributions.

\begin{figure}
\centering
\psfig{file=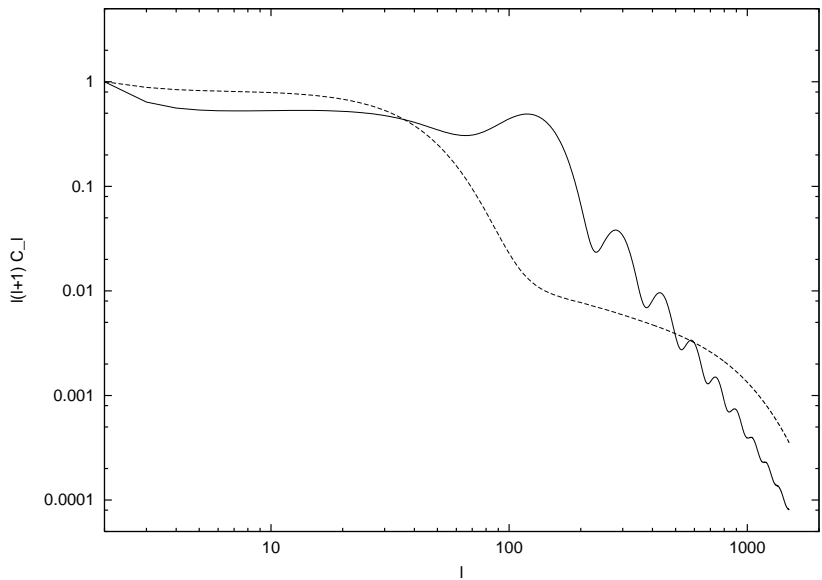,width=3.1in,angle=-90}
\caption{The vector (dashed line) and the tensor (solid line)
contributions to the CMB temperature anisotropies in a model where
${\mathcal P}={\mathcal Q}=\exp(-u^2)$}
\label{fig5}
\end{figure}

\subsection{Analytic estimate of the $C^{V,T}_\ell$ at small $\ell$}
\label{ssec_est}

As we have seen, only the integrated Sachs-Wolfe effect contributes to
the vector and tensor parts of the $C_\ell$. In this section, we
estimate them at small $\ell$. We first assume that the last
scattering surface is deep enough in the matter dominated era to
approximate ${\mathcal H}$ and $u$ by~: ${\mathcal H}=2/\eta$ and
$u=k\eta/2$. The equations (\ref{poisson_vec}) and (\ref{pert_tens}) for
$\bar\Phi_i$ and $\bar E_{kl}$ can then be rewritten [using eqns
(\ref{def_cons_vi},\ref{corr_coh_vv}-\ref{corr_coh_pit}) and
(\ref{res_coh_vec}-\ref{res_coh_tens})] as
\begin{eqnarray}
\tilde{\bar\Phi}'+{4\over\eta}\tilde{\bar\Phi} & = & 
\kappa{k\eta^{3/2}\over\sqrt2}f_6(u), \\
\tilde{\bar E}''+{4\over\eta}\tilde{\bar E}' +k^2 \tilde{\bar E} & = & 
\kappa{k\eta^{3/2}\over2\sqrt2} f_7(u),
\end{eqnarray}
the solutions of which are given by
\begin{eqnarray}
\label{est_phi_v}
\tilde{\bar\Phi} & = & 
  {\alpha\over u^4}+{k^{-3/2}\over u^4}\int_0^u v^{11/2} f_6(v)dv, \\
\label{est_e_ij}
\tilde{\bar E}   & = &  \alpha{j_1(u)\over u} 
                       +\beta{n_1(u)\over u}{k^{-3/2}\over 2u}\times\\
\nonumber&&
                       \int_0^u v^{3/2} {j_1(u)n_1(v)-j_1(v)n_1(u)\over
                                         j_1(v)n_0(v)-j_0(v)n_1(v)} f_7(v)dv.
\end{eqnarray}

We now make the ansatz that the defects disappear as soon as they
enter the horizon \IE{} that $f_6(v) \propto f_7(v) \propto Y(v-1)$
where $Y$ is the Heavyside function.

\begin{figure}
\centering
\psfig{file=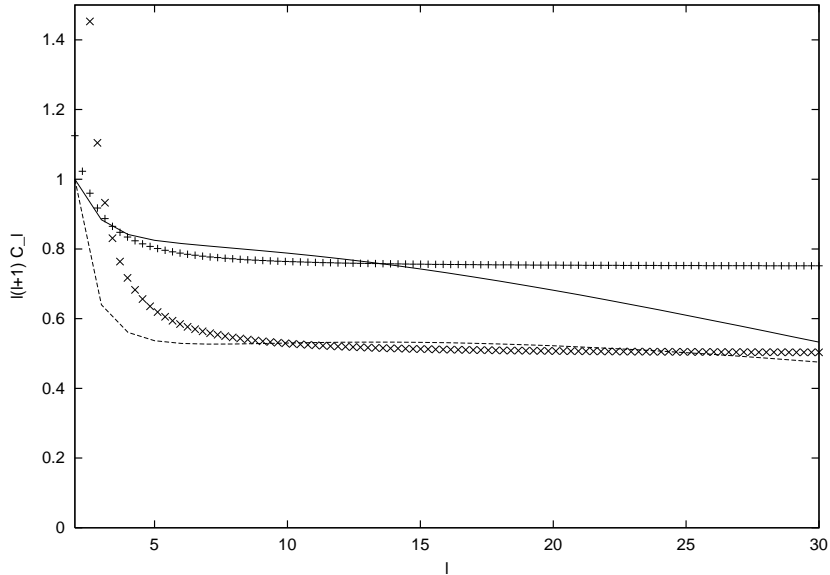,width=3.1in,angle=-90}
\caption{Comparison between the vector and tensor part to the CMB
anisotropies (solid and dotted lines respectively) with their
analytical analytic estimates ($+$-es and $\times$-es respectively)
computed in \S \ref{ssec_est}. It can be seen that the agreement is
good up to a multipole of $\ell\sim 20$ for the vector part and $\ell>
30$ for the tensor part.}
\label{fig6}
\end{figure}

Computing (\ref{est_phi_v}-\ref{est_e_ij}), their time derivative and
inserting the result in \EQNS{\ref{c_ell_v},\ref{c_ell_t}}, we obtain
that
\begin{eqnarray}
\label{est_c_ell_v}
C^V_\ell & \propto & 
  \ell(\ell+1)\int_0^A {j_\ell^2(y)\over y^2}{1\over{y+1}}\ddd y, \\
\label{est_c_ell_t}
C^T_\ell & \propto & 
  {{(\ell+2)!}\over{(\ell-2)!}}\int_0^A {j_\ell^2(y)
\over y^4}{1\over{y+1}}\ddd y,
\end{eqnarray}
with $A\equiv\eta_0/\eta_{d}-1$.  The main contribution to the two
integrals comes from the points $y\simeq l$ (since the Bessel
functions are peaked around that point). In a matter dominated
universe, we have that $A=\sqrt{a_0/a_d}-1\simeq32$ and for $l<30$ we
will assume that $A\sim\infty$. With that approximation, we are led to
compute the integrals
\begin{equation}
{\mathcal B}^\ell_p\equiv\int_0^\infty {j_\ell^2(y)
\over y^{p-1}}{1\over{y+1}}\ddd y \qquad -1<p<2\ell+2,
\end{equation}
with $p=5$ and $p=3$ for the tensor and vector modes respectively.
Using the relation \cite{gradstein}
\begin{eqnarray}
{\mathcal A}^\ell_p&\equiv&\int_0^\infty {j_\ell^2(y)\over y^{p}}\ddd y\\
\nonumber
&=&{\pi\over2^{2+p}}{\Gamma(1+p)\over\Gamma^2(1+p/2)}
{\Gamma(\ell+1/2-p/2)\over\Gamma(\ell+3/2+p/2)}\quad,\quad -1<p<2\ell+1,
\end{eqnarray}
it can be shown that
\begin{equation}
{\mathcal B}^\ell_p={\mathcal A}^\ell_p(1-{\mathcal R}^\ell_p)
\end{equation}
with
\begin{eqnarray}
0 & \leq{\mathcal R}^\ell_5\leq & {15\pi\over3.2^{11}}{1\over(\ell-5/2)}, \\
0 & \leq{\mathcal R}^\ell_3\leq & {9\pi\over2^{5}}{1\over(\ell-3/2)}.
\end{eqnarray}
Thus, the $\ell$ dependance of $C^{V,T}_\ell$ at small $\ell$ is given by
\begin{eqnarray}
\label{est_c_ell_v2}
\ell(\ell+1)C^V_\ell&\propto&{\ell(\ell+1)\over(\ell+2)(\ell-1)},\\
\label{est_c_ell_t2}
\ell(\ell+1)C^T_\ell&\propto&{\ell(\ell+1)\over(\ell+3)(\ell-2)}
\quad,\quad 3\leq \ell \leq 20.
\end{eqnarray}
This solution is compared to our numerical results on \FIG{\ref{fig6}}
where $C^T_\ell/C^V_\ell$ and its analytic estimate are plotted (when
$f_6 = f_7$). The agreement is excellent for $3\leq \ell\leq 20$ as
expected.

This estimation only assumes that we are in a matter dominated
universe, that $\eta_0/\eta_d-1<20$ and that $f_6\propto f_7 \propto
Y$. The two first approximations are well satisfied for small
multipoles. The ansatz concerning the modelisation of the functions
$f_6$ and $f_7$ will affect only the numerical pre-factors in
\EQNS{\ref{est_c_ell_v}-\ref{est_c_ell_t}} and
(\ref{est_c_ell_v2}-\ref{est_c_ell_t2}), as long as the scale of decay
of the defects is comparable with the horizon. 

\subsection{Conclusions}

In this article, we have studied the signature of coherent scaling
defects on the cosmic microwave background. For that purpose, we have
developped a semi-analytic formalism to describe the topological
defect network as well as two numerical codes (based respectively on a
fluid and a Boltzmann approach) to compute the scalar, vector and
tensor components of the coefficients $C_\ell$ of the CMB anisotropy
correlation function.

Concerning the scalar component, and contrarily to what happens in
standard inflationary scenarii, there is an important contribution
from the integrated Sachs-Wolfe term, which does not in general build
up as a plateau. In the generic case, this term dominates at low
$\ell$ so that the power on large scales is of the same order or even
higher than on small scales. Such characteristics do not fit the
present observations \cite{rev_exp}.

However, one can improve the model in various ways. First one can find
special combinations of the arbitrary functions which yield results
that are more compatible with the data (see \FIG{\ref{fig4}} and
\cite{rdxx}). But of course such a fine tuning needs to be
justified. Second one can play with the cosmological parameters.
Third one can extend our approach to more realistic defects
(incoherent defects, loss of scaling ...). Finally one can consider
defects that are produced at the end of an inflationary period
\cite{chm} (and not, as in this paper, in an up to then perfectly
homogeneous and isotropic universe). Various extensions of the
standard model of particle physics predict such a possibility (see
\EG{} \cite{rachel}).  We plan to explore those various improvements in
future work \cite{rdxx}.

\section*{Acknowledgments}

We thank David Langlois, Neil Turok and Mairi Sakellariadou for
helpful discussions.

\end{document}